\documentclass[aps,prc,showpacs,showkeys,nofootinbib]{revtex4}
\usepackage[dvips]{graphicx}
\usepackage{amsmath}
\usepackage{amssymb}

\newcommand{\az}{_{A,Z}}

\begin{document}

\title{Statistical Model for a Complete Supernova Equation of State}

\author{Matthias Hempel}
\email{m.hempel@thphys.uni-heidelberg.de}

\author{J\"urgen Schaffner-Bielich}
\email{schaffner-bielich@thphys.uni-heidelberg.de}

\affiliation{Institute for Theoretical Physics, Ruprecht-Karls-University, Philosophenweg 16, 69120 Heidelberg, Germany}

\date{\today}

\begin{abstract}
A statistical model for the equation of state and the composition of supernova matter is presented with focus on the liquid-gas phase transition of nuclear matter. It consists of an ensemble of nuclei and interacting nucleons in nuclear statistical equilibrium. A relativistic mean field model is applied for the nucleons. The masses of the nuclei are taken from nuclear structure calculations which are based on the same nuclear Lagrangian. For known nuclei experimental data is used directly. Excluded volume effects are implemented in a thermodynamic consistent way so that the transition to uniform nuclear matter at large densities can be described. Thus the model can be applied at all densities relevant for supernova simulations, i.e.~$\rho=10^5 - 10^{15}$ g/cm$^3$, and it is possible to calculate a complete supernova equation of state table. The model allows to investigate the role of shell effects, which lead to narrow-peaked distributions around the neutron magic numbers for low temperatures. At larger temperatures the distributions become broad. The significance of the statistical treatment and the nuclear distributions for the composition is shown. We find that the contribution of light clusters is very important and is only poorly represented by $\alpha$-particles alone. The results for the equation of state are systematically compared to two commonly used models for supernova matter which are based on the single nucleus approximation. Apart from the composition, in general only small differences of the different equations of state are found. The differences are most pronounced around the (low-density) liquid-gas phase transition line where the distribution of light and intermediate clusters has an important effect. Possible extensions and improvements of the model are discussed.
\end{abstract}



\maketitle

\section{Introduction}
\label{intro}

The equation of state (EOS) of hot and dense matter plays a fundamental role in the understanding of core-collapse supernovae (SNe) and plenty of other astrophysical scenarios, like e.g.~mergers of compact stars or cooling proto-neutron stars. In hydrodynamic simulations of such systems the EOS gives the crucial information about the properties of matter under these extreme conditions. So far, even the most comprehensive numerical studies of core-collapse SNe have difficulties to achieve successful explosions within the progenitor mass range $10$ M$_\odot  \leq M_{prog} \leq 15$ M$_\odot$. Explosions in spherical symmetry where accurate three-flavor Boltzmann neutrino transport can be applied, have only been obtained for an 8.8 M$_\odot$ O-Ne-Mg-core \cite{kitaura06, fischer09}. In general, multidimensional simulations are expected to achieve explosions, but they are computationally very expensive \cite{bruenn06,janka08,marek09}. In more detail, several explosion mechanisms are proposed from different groups: the neutrino-driven \cite{bethe85} the magneto-rotational \cite{leblanc70,bisno71} or the acoustic mechanisms \cite{burrows06}. In addition to multidimensional effects and the aforementioned mechanisms, an improved EOS, uncertainties in the neutrino opacities and missing nuclear effects could help to revive the shock wave and finally trigger the explosion. As an extreme example, in Ref.~\cite{sagert09} a successful explosion in spherical symmetry of a 15 M$_\odot$ progenitor star was obtained, only due to the use of an EOS which incorporates quark degrees of freedom and an appearance of a mixed phase of quarks and nucleons in the early postbounce phase. 

At present, there exist only two (hadronic) EOSs commonly used in the context of core collapse SNe. The EOS developed by Shen et al.~\cite{shen98, shen98_2} and the EOS from Lattimer and Swesty (LS) \cite{lattimer91}. The two EOSs are based on different models for the nuclear interactions. The former uses a relativistic mean field (RMF) approach, which we will also apply in our model. Nuclei are calculated in the Thomas-Fermi approximation. The latter is based upon a nonrelativistic parameterization of the nuclear interactions. Nuclei are described as a compressible liquid-drop including surface effects. Besides these two, there are plenty of EOSs which focus on particular aspects of nuclear matter but which are restricted to a certain range in temperature, asymmetry or density. In their range of validity they give a much more detailed description of the effects occuring there. The main difficulty in the construction of a complete EOS which is suitable for supernova simulations is the large domain in density $10^4$ g/cm$^{3} < \rho < 10^{15}$ g/cm$^3$, temperature $0 < T < 100$ MeV and (total) proton fraction $0< Y_p< 0.6$ which in principle has to be covered. In this broad parameter range the characteristics of nuclear matter change tremendously: from nonrelativistic to ultra-relativistic, from ideal gas behavior to highly degenerate Fermi-Dirac gas, from pure neutron matter to symmetric matter. All possible compositions appear somewhere in the extended phase diagram: uniform nuclear matter, nuclei in coexistence with free nucleons, free nucleons with the formation of light clusters, or an ideal gas mixture of different nuclei, just to mention a few possibilities. Thus one needs a rather simple but reliable model which is able to describe all the different compositions and the phenomenon connected with them. Furthermore, from a numerical point of view the calculation of the EOS table itself is also not trivial.

For uniform nuclear matter plenty of different models for the EOS exist and most of them could in principle be applied in the supernova context. So far, the nuclear EOS at large densities is not fixed and is still one of the main fields of current research in high energy physics. From the previous paragraph it becomes clear, that in the application for supernova physics the main difficulties arise below saturation density, where the liquid-gas phase transition of nuclear matter takes place. Matter becomes non-uniform, as light and/or heavy nuclear clusters (nuclei) form within the free nucleon gas. The uniform nuclear matter EOS is only one of the essential input information for the construction of an EOS which is suitable for the description of all possible conditions which typically occur in a core-collapse supernova. For the non-uniform nuclear matter phase further model assumptions are necessary. In this article we will choose one particular EOS for uniform nuclear matter and then will mainly focus on the modeling and the resulting properties of matter below saturation density. 

We want to emphasize that this low-density part of the EOS plays a special role in core-collapse supernovae: First of all, most of the computational time is spent in this low-density regime. Furthermore, the neutrinospheres are located at densities $\sim 10^{11}$ g/cm$^3$. The neutrino spectra, which belong to the most important observables of a supernova, are formed here and thus carry the information about the properties of low density nuclear matter. As mentioned earlier, most supernova simulations fail to achieve a successful explosion. The shock front which is initially traveling outwards after the bounce looses energy due to nuclear dissociation and neutrino emission and finally stalls at densities $\sim 10^{9}$ g/cm$^3$, see e.g.~\cite{burrows85,thompson03}. In the neutrino reheating paradigm the high energetic neutrinos of the deeper layers heat the matter between the neutrinospheres and the stalled shock so that the shock is reenergized until it finally drives of the envelope of the star. Thus the EOS of subsaturation matter is of particular importance for the possible reviving of the shock wave. 

The two EOSs mentioned before have been successfully applied in astrophysical simulations since many years. However, both of them are based on the single nucleus approximation (SNA) assuming that the whole distribution of different nuclei which forms at finite temperature can be represented by only one single nucleus. The single nucleus is found by a minimization procedure of the thermodynamic potential. The SNA has to be seen as a necessary assumption for any microscopic calculation based on one single unit cell with periodic boundary conditions, as e.g.~in Ref.~\cite{bonche81,newton09}, too. In such microscopic calculations the nuclei are formed out of the nucleons which are placed in the unit cell just by the nuclear interactions, which is a very convenient aspect of these models. Already in Ref.~\cite{burrows84} it was shown, that under most conditions the EOS is almost not affected by the SNA. But the SNA gives the composition only in an averaged way, as the spread in the distribution of nuclei can be large. In microscopic calculations quite often several similar minima of the thermodynamic potential are found, also indicating the occurrence of mixtures of different nuclei, in contrast to the SNA, see e.g.~\cite{bonche81,newton09}. Furthermore, in Ref.~\cite{souza08} it was presented for some particular EOS models, that the SNA leads to systematically larger nuclei, which was already shown in Ref.~\cite{burrows84}, too. The distribution of nuclei can influence the supernova-dynamics, as e.g.~electron capture rates are modified. The electron capture rates are highly sensitive to the nuclear composition and the nuclear structure, see e.g.~\cite{langanke00,langanke03,hix03,martinez06}. More closely connected to the present work, in Ref.~\cite{caballero06} the authors investigated models which are based on a distribution of various nuclei, so called nuclear statistical equilibrium (NSE) models. It was shown with classical molecular dynamics simulations that these models give systematically larger neutrino cross sections, leading to shorter neutrino mean free paths. In their study also the importance of the remaining uncertainty regarding the composition was pointed out. We want to note that the previously mentioned systematic change in the composition within the SNA was not analyzed in Ref.~\cite{caballero06}. The NSE distribution of nuclei was only compared to a SNA system with a nucleus with the mass and charge of the average of the distribution. The results of Ref.~\cite{sawyer05} point in the same direction: it was shown that the proper treatment of a multicomponent plasma leads to greatly reduced ion-ion correlations and thus to increased neutrino opacities.

There are further limitations of the previously mentioned EOSs, \cite{shen98, shen98_2} and \cite{lattimer91}. Both of them do not include any nuclear shell effects. A good description of nuclei is only achieved on an averaged basis. It might be seen as part of the SNA not to attribute any certain shell structure to the single nucleus which only represents the average of the whole distribution of nuclei. But in cases where only very few different or even only one kind of nuclei appears (e.g.~at low densities and temperatures), shell effects are important and should definitely be taken into account in a self-consistent way in the composition and in the EOS. Shell effects can substantially alter the composition and are crucial for the evaluation of the weak reaction rates with neutrinos and electrons. In our model we do not want to use the liquid-drop formulation or the Thomas-Fermi approximation, but rather implement as much information gained from experiments about the nuclear masses as possible. Thus our approach is very contrary to the two other EOSs, in which the nuclear interactions are the only input information required in the physical model for non-uniform nuclear matter. 

In some supernova EOSs, and also in the LS and in the Shen EOS, the distribution of light clusters is represented only by $\alpha$-particles, in the same way as the distribution of heavy nuclei is represented by only one heavy single nucleus. In the most simple case the $\alpha$-particles are described as a nonrelativistic, classical ideal gas, without any interactions with the surrounding nucleons. In contrast to this very simplified treatment, there are studies which focus exclusively on the role of light nuclei in supernovae and the medium effects connected with them. A model-independent description of low-density matter is given by the virial equation of state for a gas of light clusters with mass number $A\leq 4$, \cite{horowitz06a,horowitz06b,oconnor07}. In Ref.~\cite{arcones08} the composition of this model was compared to the composition of a simple NSE model. Up to densities $\rho\sim10^{13}$ only small differences were found, mainly an increased $\alpha$-particle fraction due to attractive nucleon-alpha interactions in the virial EOS. The more elaborated models of Refs.~\cite{sumiyoshi08, roepke09} are based on a quantum statistical approach. Effects of the correlated medium such as Pauli blocking, Bose enhancement and self-energy are taken into account, leading e.g.~to the merging of bound states with the continuum of scattering states with increasing density (Mott effect). In both works \cite{sumiyoshi08, roepke09} light clusters up to $A=4$ are considered and in Ref.~\cite{sumiyoshi08} some heavy clusters in addition. In Ref.~\cite{typel09} the quantum statistical approach is compared to a generalized RMF model which includes the light clusters as additional degrees of freedoms. In the generalized RMF model the Pauli-blocking shift is implemented in an empirical form which is derived from the quantum statistical model. In all these studies it is found that light clusters in addition to $\alpha$-particles contribute significantly to the composition, with a particular role of the deuterons. This is also one of the results of Ref.~\cite{heckel09}, where all stable nuclei with $A\leq13$ are included, but medium effects are only considered on a simplified level. The inclusion of the additional degrees of freedom of the light clusters can affect the supernova-dynamics. E.g.~in Ref.~\cite{arcones08} the influence of light nuclei on neutrino-driven supernova outflows was studied and a significant change in the energy of the emitted antineutrinos was found. As another example, in Ref.~\cite{oconnor07} it was shown, that mass-three nuclei contribute significantly to the neutrino energy loss for $T\geq4$ MeV. A different topic connected to light clusters is the possible formation of Bose-Einstein condensates, see e.g.~\cite{funaki08arxiv,funaki08,heckel09}.

A lot of knowledge about the properties of hot compressed nuclear matter was gained from statistical multifragmentation models (SMM) which are used to analyze low-energy heavy-ion collisions \cite{gross90, bondorf95}. In Refs.~\cite{botvina04,botvina05,mishustin08, botvina08} it was pointed out that the state reached in these experiments ($T \sim 3-6$ MeV, $\rho \sim 0.1 \rho_0$, with the saturation density $\rho_0$) is very similar to the conditions in a core-collapse supernova in the region between the proto-neutron star and the shock front. It is very attractive that the same well-established models which are used to describe matter in terrestrial experiments can be applied for matter in some of the most energetic explosions in the universe. 

In an astrophysical context these models are usually called NSE models. In this chemical picture the bound states of nucleons are treated as new species of quasi-particles. NSE models are in principle extended Saha-equations, as presented in Ref.~\cite{clifford65}. Within this approach the whole distribution of light and heavy nuclei can be included by construction. Furthermore, it is very easy to incorporate experimentally measured masses. This classical approach gives an excellent description as long as matter is sufficiently dilute that the nuclear interactions are negligible and if the temperature is so low, that the structure of the nuclei is not changed significantly, see e.g.~\cite{ignatiuk78}. If the whole distribution of nuclei is taken into account, it becomes rather difficult to implement a proper description of the medium effects on the nuclei, which become important at large densities. Thus, especially the transition to uniform nuclear matter which happens around $1/2 \rho_0$ leads to difficulties for NSE models. Obviously, in the high density regime close to saturation density, more microscopic SNA-models give a more reliable description. In this high density regime also very exotic nuclear structures, commonly called the "nuclear pasta" phases, appear \cite{sonoda08}. In a recent 3D Skyrme-Hartree-Fock calculation \cite{newton09} it was shown, that this frustrated state of nuclear matter is characterized by a large number of local energy minima, for different mass numbers and for different nuclear configurations. Thus one can expect that many different pasta shapes will coexist at a given temperature and density, which would require a statistical treatment beyond the present capabilities.

In addition to the subtleties at large densities, the assumption of NSE itself, in the sense that chemical equilibrium is established, is only valid for temperatures larger than $\sim$ 0.5 MeV. For lower temperatures, the nuclear reactions are too slow compared to the typical dynamical timescales of ms within a supernova. Anyhow we will present results below $T=0.5$ MeV for completeness and for the sake of comparison with larger temperatures. For these low temperatures one has to keep in mind that they do not represent the actual conditions in a SN, but rather the groundstate of matter after a sufficiently long time. 

Before presenting our own model, we want to discuss shortly the characteristics of some of the existing NSE models. In Ref.~\cite{ishizuka03} Ishizuka et al.~included 9000 nuclei from the theoretical mass table of Myers and Swiatecki \cite{myers90}. Excited states of the nuclei are treated with an internal partition function in the same manner as we will do. However, the model of Ishizuka et al.~does not consider any nuclear interactions or phenomenological excluded volume corrections. In this sense it is a rather pure NSE model. The SMM model of Botvina and Mishustin \cite{botvina04,botvina05,mishustin08,botvina08} does not use any tabulated binding energies, but is based on a liquid-drop parameterization of the nuclear masses. The parameters of the SMM are extracted from nuclear phenomenology and provide a good description of multifragmentation data. Shell effects are not included, but in fact they are expected to be weak at the large temperatures for which this model is designed for. On the other hand the phenomenological SMM allows to include arbitrary heavy nuclei and is easy to modify to explore certain aspects of the nuclear interactions. E.g.~nuclear excited states do not have to be taken into account explicitly, but temperature effects are included in a temperature dependent part of the nuclear binding energies, with a separate volume and surface contribution. The SMM also contains an excluded volume correction, which is equivalent to our formulation at low densities. The NSE model of Blinnikov et al.~\cite{blinnikov09} uses the tabulated theoretical nuclear binding energies of Ref.~\cite{koura05}. In their model interactions of the free nucleons as well as the modification of the nuclear surface energy due to the presence of the free nucleons is taken into account. Nuclear excited states are described by the partition functions of Engelbrecht \cite{engelbrecht91}. Besides the use of different nuclear interactions and partition functions, the major difference of this model compared to ours, which will be presented below, is, that excluded volume effects have not been implemented. In all the NSE models discussed above, the Coulomb energies are included in the Wigner-Seitz approximation, which we will also use in our model.

Despite the principle problems NSE models have to cope with at large densities, we want to construct a NSE model which allows to bridge the critical region from some percent of saturation density up to uniform nuclear matter. By using a NSE model we keep the rather simple but accurate description of low-density and low-temperature matter. For uniform nuclear matter a well-known RMF EOS will be applied. Our new NSE model shall be able to give a reasonable description of the transition to uniform nuclear matter, which in the microscopic SNA models is achieved automatically. For that we include the nuclear interactions also in the unbound nucleon contribution below saturation density. In most cases these interactions are not important, but they become crucial where the free nucleons constitute a significant fraction and the interactions are necessary for the liquid-gas phase transition. Furthermore, we develop a thermodynamic consistent description of excluded volume effects, so that we achieve the right asymptotic behavior for very dense and very dilute nuclear matter. We are aware that we apply the NSE description at densities at which we can not control all effects of the nuclear interactions any more, and our very phenomenological description becomes questionable. Anyhow, we want to explore the limits of a NSE model and compare the arising differences to other existing EOSs. So far an NSE model has never been applied close to saturation density and the existing models are not able to describe the transition to uniform nuclear matter at all. We can discuss the whole phase diagram of SN matter within one consistent model and can address all the aspects which were mentioned above, namely the distribution of heavy nuclei, the importance of the light cluster distribution, or the role of shell effects. As will be presented below, we will apply the same model which was used for the calculation of the nuclear masses for the interactions of the free nucleons, so that all nuclear interactions (apart from excluded volume effects) are based on the same Lagrangian. 

The article is structured as follows. In Sec.~\ref{sec_model} we will present our statistical model in detail. First we describe the different components, namely the RMF model for the unbound nucleons and uniform nuclear matter and the implementation of the nuclei, their excited states and the Coulomb energies. Then we derive a thermodynamic consistent description of the whole thermodynamic system, including excluded volume corrections. At the end of Sec.~\ref{sec_model} we address the transition to uniform nuclear matter. In Sec.~\ref{sec_res} we show and discuss our results, first for the composition and then for the EOS. The results for the EOS of our NSE model are systematically compared to the Shen and LS EOS. In Sec.~\ref{sec_sum} we summarize and draw conclusions. In the last Sec.~\ref{sec_out} we will give an outlook and discuss possible extensions of the model. Throughout the paper we use natural units where $\hbar=c=k_B=1$.

\section{Description of the model}
\label{sec_model}
In our model matter consists of nuclei, nucleons, electrons, positrons and photons. In many astrophysical scenarios neutrinos are not in chemical equilibrium with the rest of the matter and weak equilibrium is not established. Therefore we calculate the EOS without including the neutrino contribution. In this case the state of electrical charge neutral matter is well defined by the three independent variables $(T,n_B,Y_p)$, with the baryon number density $n_B$. In Ref.~\cite{hempel09} it was shown that also in the vicinity of a phase transition the neutrinos do not influence the non-neutrino EOS and thus can be added separately. Electrons are assumed to be distributed uniformly and are described as a general Fermi-Dirac gas, including the positron contribution. All Fermi-Dirac integrals for the electrons as well as for the nucleons were calculated using the very accurate and fast routines from Refs.~\cite{aparicio98, gong01}. Thus the possbible degeneracy of the nucleons is fully taken into account. The photon contribution (Stefan-Boltzmann law) is also included in the EOS. The nontrivial part of the model is the description of the baryons, as they are not distributed uniformly and their interactions are significant. For simplicity, for temperatures above or equal 20 MeV matter is assumed to be uniform, i.e. without the presence of nuclear clusters.

\subsection{Nucleons}
\label{ss_nuc}
For the unbound interacting nucleons (neutrons and protons) a RMF model is applied. Its Lagrangian is based on the exchange of the isoscalar scalar $\sigma$-, the isoscalar vector $\omega$- and the isovector-vector $\rho$-mesons:
\begin{eqnarray}
{\cal L} & = & \bar{\psi}\left[i\gamma_{\mu}\partial^{\mu} -M
-g_{\sigma}\sigma-g_{\omega}\gamma_{\mu}\omega^{\mu}
-g_{\rho}\gamma_{\mu}\tau_a\rho^{a\mu}\right]\psi \nonumber \\
 && +\frac{1}{2}\partial_{\mu}\sigma\partial^{\mu}\sigma
-\frac{1}{2}m^2_{\sigma}\sigma^2-\frac{1}{3}g_{2}\sigma^{3}
-\frac{1}{4}g_{3}\sigma^{4} \\ \nonumber
 && -\frac{1}{4}W_{\mu\nu}W^{\mu\nu}
+\frac{1}{2}m^2_{\omega}\omega_{\mu}\omega^{\mu}
+\frac{1}{4}c_{3}\left(\omega_{\mu}\omega^{\mu}\right)^2   \\ \nonumber
 && -\frac{1}{4}R^a_{\mu\nu}R^{a\mu\nu}
+\frac{1}{2}m^2_{\rho}\rho^a_{\mu}\rho^{a\mu} ,
\end{eqnarray}
where
\begin{eqnarray}
W^{\mu\nu} & = & \partial^{\mu}\omega^{\nu}
               - \partial^{\nu}\omega^{\mu} \; ,     \\ \nonumber
R^{a\mu\nu} & = & \partial^{\mu}\rho^{a\nu}- \partial^{\nu}\rho^{a\mu}
 +g_{\rho}\epsilon^{abc}\rho^{b\mu}\rho^{c\nu} \; .   \\ \nonumber
\end{eqnarray}
Non-linear $\sigma$ and $\omega$ terms are included to achieve a reasonable description of the properties of nuclei and of the equation of state of nuclear matter. For given nucleon densities the implicit equation of motion for the sigma-meson field needs to be solved numerically to achieve self-consistency. For very low number densities of the nucleon gas ($n_{nuc}< 10^{-5}$ fm$^{-3}$), where the interactions are negligible, the nucleons are treated as noninteracting ideal Fermi-Dirac gases for simplicity. The photons and their coupling to the nucleons are dropped at this point, because the contribution of the free photon gas is added separately to the EOS and the Coulomb energies will be discussed later.

\begin{table}
\begin{center}
\begin{tabular}{|c c c c c c c|}
\hline
 & $n_B^0$ [fm$^{-3}]$ & $E/A$ [MeV] & $K$ [MeV] & $M^*/M$ & $a_{sym}$ [MeV] & $M_{max}$ [M$_\odot$] \\
\hline
\hline
TMA & 0.147 & -16.0 & 318 & 0.635 & 30.7 & 2.0 \\
\hline
\end{tabular}
\end{center}
\caption{Nuclear matter and neutron star properties of the relativistic mean field model TMA \cite{toki95}. Listed are the saturation density and binding energy, the incompressibility, the effective mass at saturation, the symmetry energy and the maximum mass of a cold neutron star.}
\label{table_rmf}
\end{table}
The free parameters in the Lagrangian, the masses of the nucleons and the mesons and their coupling strengths, have to be determined by fits to experimental data. In this work we use the parameter set TMA \cite{toki95}. It is based on an interpolation of the two parameter sets TM1 and TM2 \cite{Suga94}, which were fitted to binding energies and charge radii of light (TM2) and heavy nuclei (TM1). The coupling parameters of the set TMA are chosen to be mass-number dependent to have a good description of nuclei over the entire range of mass number. For uniform nuclear matter the couplings become constants. 

Table \ref{table_rmf} lists some characteristic properties of the EOS of uniform nuclear matter, namely the saturation properties and the resulting maximum mass of a cold neutron star. The symmetry energy, saturation density and the binding energy of symmetric nuclear matter are well determined through the fit to ground state nuclei and lie in the usual range. The nuclear compressibility is rather high compared to the value of $K=240 \pm 20$ MeV \cite{shlomo06} or $K=248 \pm 8$ MeV \cite{piekarewicz06} deduced with theoretical models from experimental data on isoscalar giant monopole resonances (ISGMR) which probe nuclear matter slightly below saturation density. However, it is perceived in the literature that the extractation of $K$ from ISGMR data is not unambiguous as it is dependent on the density dependence of the symmetry energy of the nuclear interactions which are taken for the analysis of the data \cite{shlomo06,piekarewicz06,sharma09}. For RMF models without further constraints on the density dependence of the symmetry energy usually higher values for $K$ in the range of 250 to 270 MeV are obtained \cite{piekarewicz06}. The density dependence of the symmetry energy itself can be probed by different experimental observables, e.g.~by isospin diffusion and double neutron to proton ratios in heavy-ion collisions or the precise measurement of the neutron skin thickness of $^{208}$Pb \cite{brown00,typel01}. A recent compilation of various experimental results concerning the density dependence of the symmetry energy is given in Ref.~\cite{tsang09}. With a rather high pressure of $4.5$ MeV/fm$^3$ of pure neutron matter at saturation density TMA lies still at the border of these experimental constraints. At several times saturation density experimental flow data from high-energy heavy ion collisions can be used as a constraint for the EOS. A comparison with the analysis of Ref.~\cite{danielewicz06} shows a good agreement: TMA is lieing almost completely in the compatible range for an asymmetric stiff EOS. For TMA the maximum mass of a cold deleptonized neutron star, $M_{max}=2.0$ M$_\odot$, is well above the largest precisely known mass of $1.67 \pm 0.01$ M$_\odot$ of PSR J1903+0327 \cite{freire09}. Our choice of the parameter set TMA was mainly due to the availability of a sufficiently large and precise mass table (see following subsection). Because of consistency we did not want to combine one theoretical mass table with a different model for the nuclear interactions. Further details about the parameter set TMA are described in Refs.~\cite{toki95,geng05}, general reviews about the relativistic mean field model are given in Refs.~\cite{reinhard89}.

\begin{figure}
\includegraphics{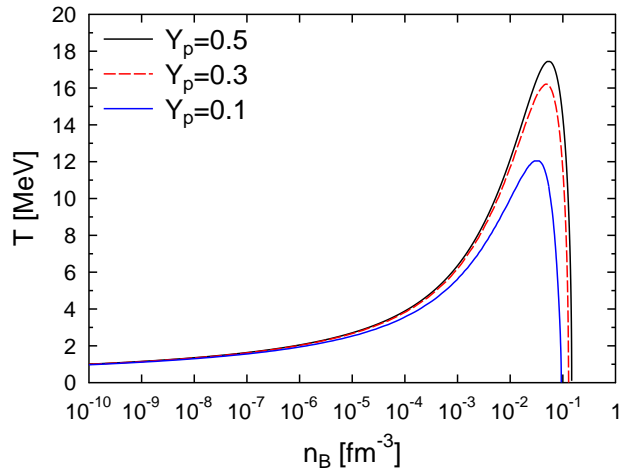}
\caption{\label{fig_bulkpd}The phase diagram of bulk nuclear matter calculated with the RMF model TMA \cite{toki95}. The lines show the binodals for different $Y_p$, at which the the liquid-gas phase transition of nuclear matter sets in and a mixed phase appears. (color version online)}
\end{figure}
Fig.~\ref{fig_bulkpd} shows the phase diagram of bulk nuclear matter calculated with the RMF model. For symmetric nuclear matter a critical temperature of 17.4 MeV is found. At $T=0$ MeV the mixed phase ends between $0.63 n_B^0$ and $n_B^0$ for the shown values of $Y_p$. At larger densities matter is uniform again. The phase diagram depends stronly on $Y_p$: For lower proton fractions the mixed phase region shrinks considerably. We will use this phase diagram for comparison with the NSE model later.

\subsection{Nuclei}
In our approach we will preferably use experimentally measured masses for the description of nuclei (mass number $A\geq2$). We take the nuclear data from the atomic mass table 2003 from Audi, Wapstra, and Thibault (AWT) \cite{AudiWapstra} whenever possible. It is very convenient that we directly can use experimental data for the construction of the EOS. We do not take any estimated, non-experimental data of the atomic mass table into account. For nuclei with experimentally unknown mass we use the results of the nuclear structure calculation \cite{geng05} in form of a nuclear mass table. This mass table is based on the same relativistic mean field model TMA which was presented in the previous section. It lists 6969 even-even, even-odd and odd-odd nuclei, extending from $^{16}$O to $^{331}100$ from slightly above the proton to slightly below the neutron drip line. The nuclear binding energies were calculated under consideration of axial deformations and the pairing was included with a BCS-type $\delta$-force. With these detailed calculations of the nuclear masses a good agreement with the experimental masses is achieved, with a rms deviation $\sigma \sim 2.1$ MeV \cite{geng05}.

\begin{figure}
\includegraphics{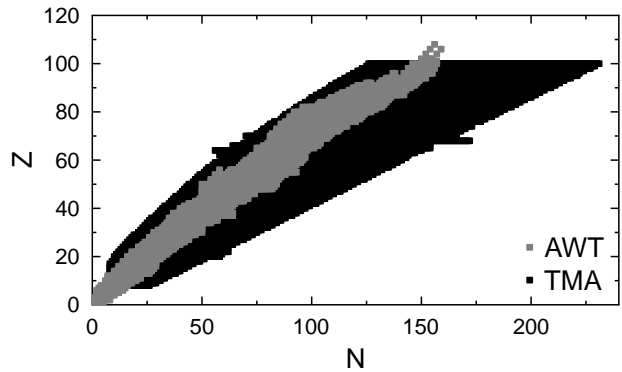}
\caption{\label{fig_nuclidechart} The proton number $Z$ and neutron number $N$ of the nuclei taken from the mass table calculated with TMA \cite{toki95,geng05} and from the experimental AWT mass table \cite{AudiWapstra}.}
\end{figure}
As long as a mixed phase of free nucleons and nuclei is favored instead of uniform nuclear matter, our statistical description includes all nuclei which are listed in the experimental AWT table or in the theoretical mass table respectively. Fig.~\ref{fig_nuclidechart} shows all the considered nuclei in a nuclear chart.

At finite temperature excited states of the nuclei will be populated and consequently the number density of a certain nucleus $(A,Z)$ with mass number $A$ and proton number $Z$ will increase. Instead of including all excited states explicitly we use a temperature dependent degeneracy function $g_{AZ}(T)$. It represents the sum over all excited states of a hot nucleus. We choose the simple semi-empirical expression for $g_{AZ}(T)$ from Ref.~\cite{fai82}:
\begin{equation}
g_{AZ}(T)=g_{AZ}^0 +\frac{c_1}{A^{5/3}}\int_0^\infty dE^* e^{-E^*/T}\exp\left(\sqrt{2 a(A) E^*}\right) \\ \nonumber
\end{equation}
\begin{equation}
a(A)=\frac A 8 (1-c_2A^{-1/3}) \rm {~MeV} ^{-1} \\ \nonumber
\end{equation}
\begin{equation}
c_1=0.2 \rm {~MeV} ^{-1},\;\; c_2=0.8 \; ,
\end{equation}
with $g_{AZ}^0$ denoting the degeneracy of the ground-state. $g_{AZ}^0$ is very small compared to $g_{AZ}(T)$ and therefore we take $g_{AZ}^0=1$ for even and $g_{AZ}^0=3$ for odd $A$ for simplicity. In this way the only information needed about the nuclei are their ground state masses.

\subsection{Coulomb energies}
\label{ss_coul}
For the calculation of the Coulomb energies, which play an important role in the determination of the composition, we assume spherical Wigner-Seitz (WS) cells for every nucleus. For an uniform electron distribution with electrons present inside and outside of a nucleus $(A,Z)$ one gets a simple expression for the Coulomb energy of the WS cell:
\begin{equation}
 E^{Coul}\az=- \frac35 \frac{Z^2 \alpha}{R(A)}\left(\frac32 x- \frac12 x^3\right)
\label{eq_ecoul}
\end{equation}
\begin{equation}
 x=\left(\frac{n_e}{n_B^0}\frac{A}{Z}\right)^{1/3} \; , \label{eq_x}
\end{equation}
where we treated the nuclei as homogeneous spheres with radius $R(A)$ of nucleons at saturation density $n_B^0$: $R(A)=(3 A/4\pi  n_B^0)^{1/3}$. $n_e$ is the electron number density, which is fixed by charge neutrality: $n_e=Y_p n_B$. The first term in the brackets corresponds to the Coulomb energy of a point-like nucleus with charge $Z$ within the electron gas. The second term in the brackets arises due to the finite size of the nucleus, with electrons located inside the nucleus. 

We do not include the Coulomb energy of the protons because of the following reasons: In principle they could be added within the WS approximation in the same way as for the nuclei, without any further complications. But first of all protons are rather light particles, so that the WS approximation and the above expression for the Coulomb energy would not be very adequate. Next and more important in our context, even when uniform charge neutral nuclear matter was reached, the Coulomb energy within the WS approximation would not vanish but would lead to a spurious contribution to the nuclear binding energy. Instead of the WS approximation one could treat the protons as an uniform background, which would screen the charge of the electrons and interact with the charge of the nuclei. Then the Coulomb energy would vanish for uniform nuclear matter as it has to be. But within our description of the thermodynamic system (which will be presented next) this would lead to numerical complications as an additional implicit equation had to be solved. Thus for simplicity we neglect the anyhow small Coulomb energy of the protons. Then the correct limit of vanishing Coulomb energy for uniform nuclear matter is also achieved.

\subsection{Thermodynamic model}
\label{ss_therm}
In our description we distinguish between nuclei and the surrounding interacting nucleons, and we still have to specify how the system is composed of the different particles. For nuclei we will apply the following classical description: All baryons (nucleons in nuclei or unbound nucleons) are treated as hard spheres with the volume $1/n_B^0$ so that the nuclei are uniform hard spheres at saturation density of volume $A/n_B^0$. Next, a nucleus must not overlap with any other baryon (nuclei or unbound nucleons). Thus the volume in which the nuclei can move is not the total volume of the system, but only the volume which is not filled by baryons. For the unbound nucleons we use a different description, because the interactions among them are already included in the RMF model. For unbound nucleons we only assume that they are not allowed to be situated inside the nuclei. We will discuss these two different excluded volume corrections in more detail later.

To derive all relevant thermodynamic quantities like e.g.~the energy density or the pressure for given $(T,n_B,Y_p)$, we start from the total canonical partition function of the system. To do that we will first consider that the entire set $\{N_i\}$ of all the particle numbers of electrons $N_e$, neutrons $N_n$, protons $N_p$ and all nuclei $\{N\az\}$ is fixed (the trivial photon contribution is taken out of the following derivation). In our model the total energy can be split into the contribution of electrons, nucleons, nuclei and the Coulomb energy. Thus the total canonical partition function is given by the product of the partition functions of the four different contributions:
\begin{equation}
Z(T,V,\{N_i\})=Z_e~Z_{nuc}~\prod_{A,Z}Z\az~Z_{Coul} \; ,
\end{equation}
with $V$ denoting the volume of the system. From the canonical partition function the canonical thermodynamic potential follows, which is the Helmholtz free energy:
\begin{eqnarray}
F(T,V,\{N_i\})&=&-T \mathrm{ln} Z \\
&=& F_e+F_{nuc}+\sum\az F\az+F_{Coul} \; .
\end{eqnarray}
In the following the different contributions to the free energy will be discussed separately in detail.

The free energy of the electrons $F_e=-T \mathrm{ln} Z_e$ is given by a general noninteracting ideal Fermi-Dirac gas, including antiparticle contributions:
\begin{eqnarray}
F_e(T,V,\{N_i\})&=&F_e^0(T,V,N_e) \; .
\end{eqnarray}
The electrons are distributed over the entire volume and are not influenced by the excluded volume effects. Then the Coulomb free energy has the following simple form:
\begin{eqnarray}
F_{Coul}(T,V,\{N_i\})&=&F_{Coul}(T,V,N_e,\{N\az\}) \nonumber \\
&=& \sum \az N\az E^{Coul}\az(V,N_e)\; . \label{eq_fcoul}
\end{eqnarray}
From Eq.~(\ref{eq_ecoul}) and (\ref{eq_x}) it is clear that the Coulomb free energy actually depends only on the number density of the electrons, $n_e=N_e/V$, which is fixed by charge neutrality, $n_e=Y_p n_B$, and the numbers of protons $N_p$ and nuclei $\{N\az\}$ but not on the volume. Thus the Coulomb free energy is also not modified by the excluded volume corrections.

The volume available to the nucleons is reduced by the volume which is filled by nuclei. Therefore the free energy of the nucleons $F_{nuc}=-T \mathrm{ln} Z_{nuc}$ is the free energy calculated with the unmodified RMF model $F^0_{nuc}$ for the available volume $V'$ which is not filled by by the volumes of the nuclei:
\begin{eqnarray}
F_{nuc}(T,V,\{N_i\})&=&F_{nuc}^0(T,V',N_n,N_p) \\
V'&=&V-\sum\az N\az V\az \\
V\az&=&AV_0=A/n_B^0 \; . \label{eq_vaz}
\end{eqnarray}
If no nuclei are present we arrive at the unmodified RMF description, as it should be.

Nuclei are treated as nonrelativistic classical particles. Also for the nuclei an excluded volume correction is introduced, but one which has a different character than the one for nucleons. The nuclei are allowed to be everywhere in the system as long as they do not overlap with any other baryon (nucleons inside of the other nuclei or the free nucleons). Regarding the effect on the nuclei, the same volume as the one of nucleons in nuclei is attributed to the free nucleons, so that according to Eq.~(\ref{eq_vaz}):
\begin{eqnarray}
V_n&=&V_p=V_0=1/n_B^0 \;.
\end{eqnarray}
Every baryon in the system reduces the free volume $\bar V$ in which the nuclei can move:
\begin{eqnarray}
\bar V&=&V-\sum\az N\az V\az -N_n V_n- N_p V_p \;.
\end{eqnarray}
Thus within our assumptions, the free energy of the nucleus $(A,Z)$ is the usual Maxwell-Boltzmann expression of a classical ideal gas in the free volume $\bar V$:
\begin{eqnarray}
F_{A,Z}(T,V,\{N_i\})&=&F\az^0(T,\bar V,N\az) \\
F\az^0&=&N\az M\az-TN\az\left(\mathrm{ln}\left(\frac{g\az(T)\bar V}{N \az}\left(\frac{M\az T}{2\pi}\right)^{3/2}\right)+1\right) \;.
\end{eqnarray}
As the volume appears only in the kinetic part of the free energy, naturally this excluded volume correction does not modify the rest mass term.

The excluded volume correction of the nuclei represents a hard-core repulsion of the nuclei at large densities close to saturation density. Instead the modification of the free energy of the unbound nucleons is purely geometric and just describes that the nucleons fill only a fraction of the total volume. It is important to note that nuclei can not be present at densities larger than saturation density within this picture, which is reasonable and wanted. This is also the reason why we chose $V_0=1/n_B^0$ for the value of the volume of a nucleon, which has to be seen as a free parameter of the model.

The volume $V$ of the system can be chosen freely and just determines the size of the system. As its value is completely arbitrary in the thermodynamic limit, a description in which all extensive quantities are replaced by their corresponding densities is more convenient. The total particle number densities are the numbers of particles per total volume:
\begin{eqnarray}
n_{n/p}&=&N_{n/p}/V \\
n\az&=&N\az/V \; .
\end{eqnarray}
For the nucleons we introduce the local number densities outside of the nuclei, given by the number of neutrons respectively protons per available volume:
\begin{eqnarray}
n'_{n/p}&=&N_{n/p}/V' \; .
\end{eqnarray}
In the following we will use these local number densities of the nucleons instead of their total number densities, as they directly set the RMF contribution of the nucleons to the EOS. After introducing the filling factor of the nucleons
\begin{eqnarray}
\xi&=&V'/V=1-\sum \az n\az V\az \label{eq_xi} \\
&=&1-\sum \az A~n\az/n_B^0
\end{eqnarray}
the total number and the electric charge density of the baryons take on the following form:
\begin{eqnarray}
n_B&=&(N_n+N_p+\sum \az A N\az)/V\\
&=& \xi(n'_n+n'_p)+\sum \az A~n\az \; ,
\end{eqnarray}
\begin{eqnarray}
n_BY_p&=&(N_p+\sum \az Z N\az)/V\\
&=& \xi n'_p+\sum \az Z n\az \; .
\end{eqnarray}
$\xi=1$ corresponds to the case when only free nucleons are present. For $\xi=0$ the nuclei fill the entire space so that there is no available volume for the free nucleons left.

To replace $\bar V$ in the expressions used later the volume fraction $\kappa$ is introduced:
\begin{eqnarray}
\kappa&=&\frac{\bar V}V \; .
\end{eqnarray}
It is the fraction of the free volume $\bar V$ in which the nuclei can move of the total volume $V$. It depends only on $n_B$:
\begin{eqnarray}
\kappa&=&1-\frac1{n_B^0V}\left(\sum \az N\az+N_n+N_p\right)\\
&=&1-n_B/n_B^0 \label{eq_kappa} \; .
\end{eqnarray}
$1-\kappa$ is the volume fraction which is filled by baryons. If $\kappa=1$ ($n_B=0$) then the free volume is equal to the total volume, for $\kappa=0$ ($n_B=n_B^0$) the entire space is filled by baryons and the free volume vanishes.

After having specified the free energy, all thermodynamic quantities can be derived consistently in the standard manner as derivatives of the free energy. Only the internal energy density $\epsilon=U/V$ has to be determined by the inverse Legendre transformation of the free energy density $f=F/V$, $\epsilon=f+Ts$, with $s=S/V$ denoting the entropy density.

In the intensive formulation the free energy density becomes:
\begin{eqnarray}
f&=&f_e^0(T,n_e)+\sum \az f\az^0(T,n\az)+f_{Coul}(n_e,\{n\az\})+\xi f_{nuc}^0(T,n'_n,n'_p)-T\sum \az n\az \mathrm{ln}(\kappa)\; , \label{eq_f}
\end{eqnarray}
\begin{eqnarray}
f_{Coul}(n_e,\{n\az\})&=&\sum \az n\az E^{Coul}_{A,Z}(n_e)\; ,\\
f\az^0(T,n\az)&=& n\az\left(M\az-T-T\mathrm{ln}\left(\frac{g\az(T)}{n \az}\left(\frac{M\az T}{2\pi}\right)^{3/2}\right)\right)\; .
\end{eqnarray}
The first two terms in eq.~(\ref{eq_f}) are the ideal gas expressions of the electrons and the nuclei. The Coulomb free energy of the nuclei appears in addition. The free energy density of the nucleons is weighted with their volume fraction in the fourth term. This can be expected as the free energy is an extensive quantity. The last term arises directly from the excluded volume corrections of the nuclei. Because of this term, as long as nuclei are present, the free energy density goes to infinity when approaching saturation density ($\kappa \rightarrow 0$). Thus uniform nuclear matter will always set in slightly before saturation density is reached.

The entropy density can be written in the following form:
\begin{eqnarray}
s&=&s_e^0(T,n_e)+\sum \az s\az^0(T,n\az)+\xi s_{nuc}^0(T,n'_n,n'_p)+ \sum \az n\az \mathrm{ln}(\kappa)\; , \label{eq_snuclei}
\end{eqnarray}
\begin{eqnarray}
s\az^0(T,n\az)&=&n\az\left(\mathrm{ln}\left(\frac{g\az(T)}{n \az}\left(\frac{M\az T}{2\pi}\right)^{3/2}\right)+\frac52+\frac{\partial g}{\partial T}\frac T g \right) \; . \label{eq_s0}
\end{eqnarray}
Analog expressions as in the free energy density appear. As the Coulomb energy is not taken to be temperature dependent it does not give a contribution to the entropy. In the ideal gas expression of the nuclei, eq.~(\ref{eq_s0}), an additional contribution from the temperature dependent degeneracy arises. The excluded volume correction term in eq.~(\ref{eq_snuclei}) expresses the reduction of the available number of states for the nuclei with increasing density.

The energy density looks similar:
\begin{eqnarray}
\epsilon&=&\epsilon_e^0(T,n_e)+\xi \epsilon_{nuc}^0(T,n'_n,n'_p)+\sum \az \epsilon\az^0(T,n\az)+f_{Coul}(n_e,\{n\az\})\; ,
\end{eqnarray}
\begin{eqnarray}
\epsilon\az^0(T,n\az)&=& n\az\left(M\az+\frac32T+\frac{\partial g}{\partial T}\frac {T^2} g\right)\; .
\end{eqnarray}

The total pressure becomes:
\begin{eqnarray}
p&=&p_e^0(T,n_e)+p_{nuc}^0(T,n'_n,n'_p)+\frac1{\kappa}\sum \az p\az^0(T,n\az)+ p_{Coul}(n_e,\{n\az\}) \label{eq_p}
\end{eqnarray}
\begin{eqnarray}
p\az^0(T,n\az)&=&Tn\az
\end{eqnarray}
where the ideal gas pressure of the nuclei is increased by $1/\kappa$ as their free volume is reduced by the excluded volume of the baryons.
$p_{Coul}$ denotes the negative Coulomb pressure:
\begin{eqnarray}
p_{Coul}(n_e,\{n\az\})&=&-\sum \az {n\az}\frac35 \frac{Z^2 \alpha}{R(A)}\left(\frac12 x-\frac12 x^3\right) \; .
\end{eqnarray}

The chemical potential of the electrons is reduced by the Coulomb interactions:
\begin{eqnarray}
\mu_e&=&\mu_e^0(T,n_e)+\frac{p_{Coul}(n_e,\{n\az\})}{n_e}\; .
\end{eqnarray}
The chemical potential of the neutrons and protons is:
\begin{eqnarray}
\mu_{n/p}&=&\mu_{n/p}^0(T,n'_n,n'_p)+\frac1{\kappa}\sum \az p\az^0(T,n\az) V_{0} \; . \label{eq_mun}
\end{eqnarray}
Because of the excluded volume corrections mechanical work has to be exerted upon the pressure of the nuclei to add an additional nucleon. The chemical potential of the nuclei encounters an even stronger modification:
\begin{eqnarray}
\mu \az&=& \mu \az ^0(T,n \az)+E^{Coul}\az(n_e)\\
&&+\left(p^0_{nuc}(T,n_n,n_p)+\frac1{\kappa}\sum \az p\az^0(T,n\az)\right)V\az-T \mathrm{ln}(\kappa) \; ,
\end{eqnarray}
\begin{eqnarray}
\mu \az ^0(T,n \az)&=&M\az-T \mathrm{ln}\left(\frac{g\az(T)}{n \az}\left(\frac{M\az T}{2\pi}\right)^{3/2}\right)
\end{eqnarray}
Besides the chemical potential of an ideal gas the Coulomb energy of the nucleus appears. Furthermore to add an additional nucleus volume work has to be performed not only against the pressure of the nuclei, but also against the nucleonic pressure. The last term arises directly from the excluded volume correction and shows the increase of the chemical potential when the density becomes close to saturation density.

The presented approach for the excluded volume corrections is thermodynamically fully consistent and the part for the nuclei is equivalent to the method described in Ref.~\cite{rischke91} in which a grand-canonical formulation was used.

So far, all the particle number densities, $n_n$, $n_p$, $\{n\az\}$, $n_e$ were treated as fixed variables. In the following the equilibrium conditions for the baryons will be derived for the assumption of baryon number and proton number (or proton fraction) conservation which is equivalent to baryon number and isospin conservation. The electron contribution is fixed by charge neutrality. We note that our procedure, to derive the thermodynamic variables from the thermodynamic potential for given $n_n$, $n_p$, $\{n\az\}$, $n_e$ first, and to implement chemical equilibrium of the baryons afterwards, gives the correct result and is much simpler than doing it the other way round. The equilibrium conditions only set the baryonic composition but do not change the other thermodynamic functions. 

For given $n_B$ and $Y_p$ the internal variables  $n_n$, $n_p$ and $\{n\az\}$ are no longer independent. After including the conservation laws with the help of two Lagrange multipliers the first and second law of thermodynamics lead to the following relation which expresses the chemical equilibrium between nuclei and nucleons:
\begin{eqnarray}
\mu\az&=&(A-Z)\mu_n+Z\mu_p \; . \label{eq_chemeq}
\end{eqnarray}
With this condition only two degrees of freedom (e.g.~$n_B$ and $Y_p$) remain, which have to be specified.
Then Eqs.~(\ref{eq_mun})-(\ref{eq_chemeq}) can be combined to
\begin{eqnarray}
&&n \az=\kappa~g\az(T)\left(\frac{M\az T}{2\pi}\right)^{3/2}\exp\left(\frac{(A-Z)\mu_{n}^0+Z\mu_{p}^0-M\az-E^{Coul}\az-p^0_{nuc}V\az}T\right) \; . \label{eq_naz}
\end{eqnarray}
Because all the baryons (including the free nucleons) contribute equally to the excluded volume of the nuclei, the pressure of the nuclei drops out in the equilibrium condition Eq.~\ref{eq_chemeq} and the number density of the nuclei can be written in this form. 

We want to emphasize that $\mu_n^0$ and $\mu_p^0$ contain the RMF interactions of the nucleons. As they appear in Eq.~(\ref{eq_naz}) the interactions of the free nucleons are thus coupled to the contribution of the nuclei. Compared to the generalized RMF model of Typel et al.~\cite{typel09}, the mutual counteracting in-medium self energy and the Pauli-blocking shifts of the light clusters appear in addition in their model. Furthermore, in our approach the bound nucleons do not contribute to the source term of the meson fields. In our model the Mott effect and the dissolution of clusters at large densities is mimicked only by the excluded volume corrections.

The factors $\xi$ and $\kappa$ in eq.~(\ref{eq_naz}) themselves depend on the number densities of the free nucleons and/or nuclei, and therefore the set of equations (\ref{eq_xi})-(\ref{eq_kappa}) and (\ref{eq_naz}) still has to be solved numerically in a self-consistent way. Within this formulation the values of $n'_n$ and $n'_p$ determine the total baryon number density and the proton fraction. Thus for given $n_B$ and $Y_p$ only two nested root-findings have to be performed, in addition to the root-finding for the RMF equations for the nucleons. Thermodynamic consistency and consistency of the mass fractions is reached on a high level within the calculation and the relative error never exceeds $10^{-6}$.

\subsection{Transition to uniform nuclear matter}
\label{ss_trans}
The non-uniform nuclear matter phase which was described in the previous subsections has to be compared to uniform nuclear matter, calculated with the RMF model described in Subsec.~\ref{ss_nuc}, to determine whether it is energetically favored or not. The excluded volume effects assure that the non-uniform nuclear matter phase with nuclei will always be replaced by uniform nuclear matter before saturation density is reached. For given number densities and temperature the corresponding thermodynamic potential which has to be minimized is the Helmholtz free energy. Within this procedure the temperature, number densities and the Helmholtz free energy would change continuously across the transition to uniform nuclear matter by construction, but all other thermodynamic quantities would behave discontinuously in general. To ensure thermodynamic stability, e.g.~with respect to isothermal compression, $\partial p / \partial n_B)_{T,Y_p}>0$, thus the use of the Helmholtz free energy is not sufficient. Instead some kind of Maxwell construction is required for the description of the phase transition from the mixture of nuclei and nucleons as the first phase to uniform nuclear matter as the second.

Here we take the most simple choice: We connect non-uniform with uniform nuclear matter by requiring local charge neutrality and the same local proton fraction in both phases. We note that the same mixed phase construction was used in the LS EOS. This description of the phase transition (locally fixed $Y_p$ and local charge neutrality), all its implications and the derivation of the corresponding equilibrium conditions are given in detail in Ref.~\cite{hempel09}, denoted there by case Ic. Besides temperature and pressure equilibrium, only the baryon chemical potential in the form $\mu_B=(1-Y_p)\mu_n+Y_p(\mu_p+\mu_e)$ has to be equal in the two phases. With this choice the two phases behave like one-component, simple bodies, which means that the total pressure including electrons remains constant across the transition. By using these stringent conditions we will get discontinuities in the second derivatives of the free energy. Like in the familiar case of the liquid-gas phase transition of H$_2$O a mixed phase will form. Here the mixed phase consists of non-uniform nuclear matter (as described in the previous subsections) in coexistence with uniform nuclear matter. The two phases have the same $Y_p$ and $T$ but will have different $n_B$ in general. Like in water, this mixed phase is completely characterized by the properties of the first phase at the onset and of the second phase at the endpoint of the phase transition and the volume fractions of the two phases which is set by $n_B$. During the transition the properties of each of the two phases remain the same, it is only the volume fraction which is changing. This makes this mixed phase very easy to calculate. Furthermore another simplification is used: the point at which the pressure in the two phases is equal is determined in an approximative way. To save computational time the two phases are only compared at the density grid-points of the final EOS table. Then the two phases are connected by a thermodynamic consistent interpolation. We want to stress that the liquid-gas phase transition of nuclear matter is almost completely described by the statistical model alone. It is only the transition to uniform nuclear matter for which we need the mixed phase construction. An explanation why uniform nuclear matter is not reached automatically is given in the following. 

In the bulk approximation the following behavior for the transition to uniform matter, i.e.~close to the endpoint of the liquid-gas phase transition (see Fig.~\ref{fig_bulkpd}), is expected: Besides of the restricted parameter-space in $(T, Y_p)$ where retrograde condensation takes place, the volume fraction of the liquid phase (nuclei) grows with increasing density until this phase occupies the entire space, the gas phase disappears and uniform nuclear matter is reached, see \cite{glendenning92, muller95}. In our model the mass and charge number of the nuclei is limited by the nuclear mass table and thus the nuclear clusters are not able to grow arbitrary in size. Still it is possible that the volume fraction of nuclei approaches unity, but as the nuclei are described as unchangeable particles they are obviously not able to evolve to uniform nuclear matter, and the Maxwell construction is necessary. With this construction the non-uniform phase consisting of nuclei and nucleons is replaced successively by uniform nuclear matter with increasing density. One can interprete the second phase (uniform nuclear matter) as an infinitely large, charge neutral cluster with fixed $Y_p$ which occupies a volume fraction which increases with density. This interpretation of the denser phase in the phase transition is also used in Refs.~\cite{bugaev00,bugaev01}. In these works an analytic solution of a simplified SMM in the thermodynamic limit was studied, in which excluded volume effects were treated self-consistently. The behavior of the mixed phase in our model is qualitatively similar to their results or of e.g.~Ref.~\cite{muller95} in which bulk nuclear matter was studied. Interestingly, the results of the recent work \cite{newton09} seem to indicate, that even within a 3D Skyrme-Hartree-Fock calculation the phase transition to uniform nuclear matter requires a mixed phase construction.

\section{Results}
\label{sec_res}
\subsection{Composition}
\label{ss_comp}
\begin{figure}
\includegraphics{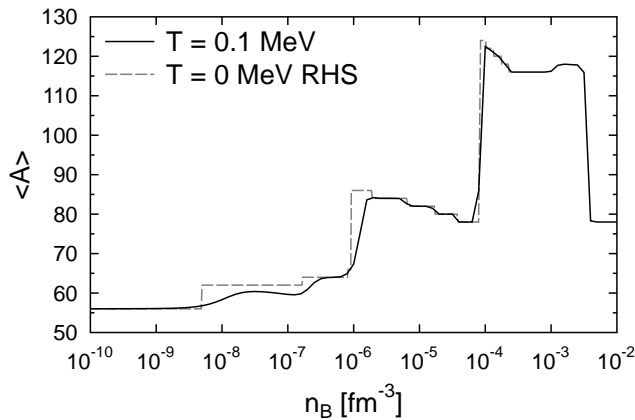}
\caption{\label{fig_abeta}The average mass number of heavy nuclei $\left<A\right>$ as a function of the baryon number density $n_B$ in $\beta$-equilibrium. The results of the present work at $T=0.1$ MeV are compared to the results from Ref.~\cite{ruester06} at temperature $T=0$ MeV, which are also based on the mass table TMA but in which the lattice energy is included explicitly.}
\end{figure}
Fig.~\ref{fig_abeta} depicts the composition in terms of the average mass number of the heavy nuclei $\left<A\right>=\sum_{A,Z\geq 6}A n\az / \sum_{A,Z\geq 6} n\az$ as a function of $n_B$ for $\beta$-equilibrated neutrino-free matter, i.e.~$\mu_n=\mu_p-\mu_e$. Here and in the following we use the proton number to differentiate between light ($Z\leq 5$) and heavy ($Z\geq 6$) nuclei. The results of the present investigation at $T=0.1$ MeV are compared to a detailed calculation of the outer crust of a neutron star at $T=0$ \cite{ruester06} which is based on the same nuclear mass table TMA. Instead of the simplified manner described in Secs.~\ref{ss_coul} and \ref{ss_therm}, in Ref.~\cite{ruester06} the Coulomb energy of a body-centred-cubic lattice is incorporated explicitly in the EOS and only the single ground-state nucleus is determined. For $T=0.1$ MeV the temperature effects are small and only lead to a smoothing of the stepwise change in the composition at $T=0$. Regarding the composition, the good agreement between the two different calculations shows that the simplified treatment of the Coulomb energies in the statistical model of the present work does not cause any significant differences. We conclude, that an excellent description of nuclear matter composition at low temperatures and densities is achieved, which incorporates shell effects.

In both calculations the system exhibits nuclei with smaller $Z/A$ for increasing density, as the electrons become relativistic and their contribution to the free energy becomes larger. The decrease in $Z/A$ leads to an in overall increasing mass number, up to $n_B \sim 10^{-4}$ fm$^{-3}$. Shell effects are strongly pronounced: for $10^{-6}$ fm$^{-3} < n_B <10^{-4}$ fm$^{-3}$ only nuclei with the magic neutron number 50 are present and around $10^{-4}$ fm$^{-3}$ only the magic neutron number 82 is populated. It is important to note that up to $\sim 3 \times 10^{-5}$ fm$^{-3}$ the composition is given entirely by nuclei whose mass is taken from experimental data. Around $n_B\sim 2.7 \times 10^{-4}$ fm$^{-3}$ the calculation of Ref.~\cite{ruester06} ends, where the so called neutron drip is reached, at which free neutrons begin to appear. In the statistical model the nucleus $^{116}$Se initially remains the favored nucleus after the neutron drip. Then $^{118}$Se appears and at larger densities $^{78}$Ca, a nucleus with very low $Z/A\sim0.26$ becomes the most abundant nucleus. Obviously, in our statistical model the composition is restricted on the nuclei which are listed in the used nuclear mass table. This could be the reason for the unexpected decrease of $<A>$ at densities close to saturation.

We note that our model is actually not very suitable for the description of neutron star matter in the inner crust, corresponding to the high density part of Fig.~\ref{fig_abeta}. In contrast to matter in SN, in neutron stars the proton fraction is very low causing a large mass fraction of unbound neutrons. Thus the interactions between free neutrons and nuclei, which change the structure of the nuclei, are very important. Later we will discuss this aspect further. We will show that for typical supernova conditions, in the regime where the contribution of the nuclei is important, the mass fraction of free nucleons is low instead.

\begin{figure}
\includegraphics{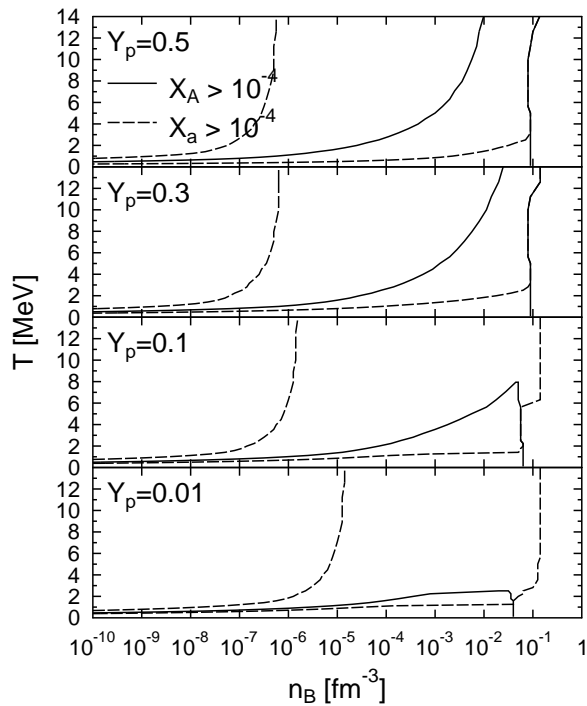}
\caption{\label{fig_pdyp}The phase diagram of nuclear matter for different proton fractions $Y_p$ in the $T-n_B$ plane. Solid (dashed) lines enclose the region where the mass fraction of heavy nuclei with $Z \geq 6$ (light nuclei with $Z \leq 5$) exceeds $10^{-4}$.}
\end{figure}
\begin{figure}
\includegraphics{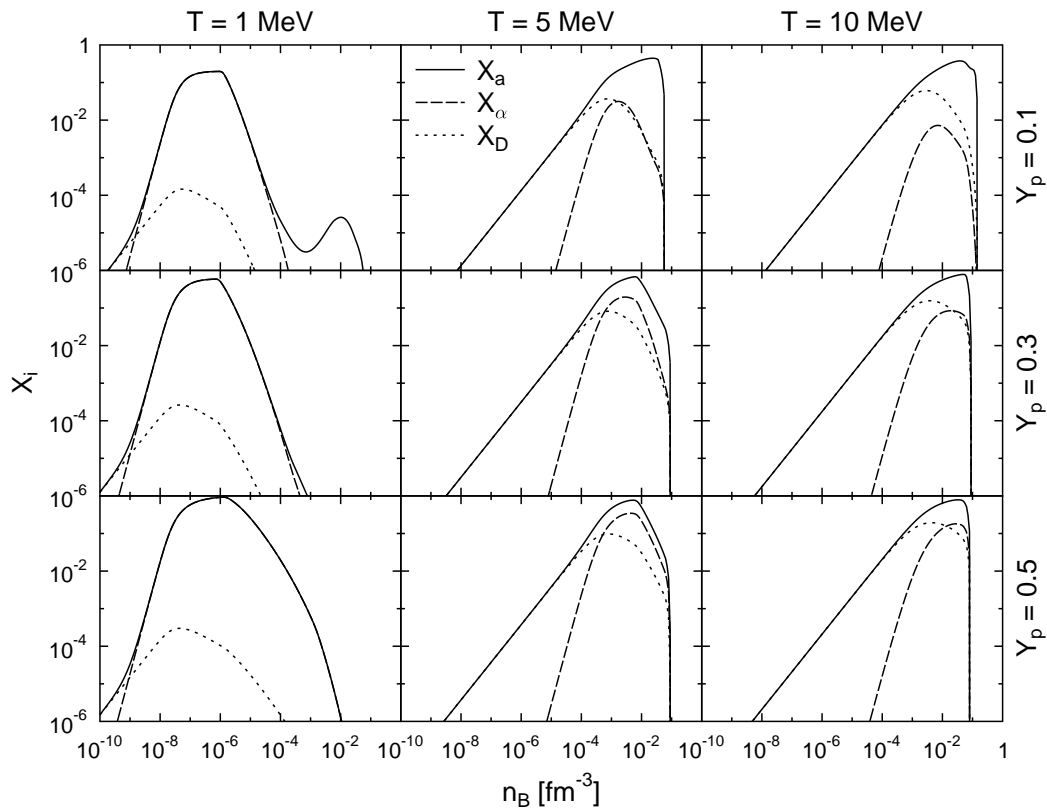}
\caption{\label{fig_xlight}The mass fractions of all light nuclei with $Z\leq 5$ $X_a$, of $\alpha$-particles $X_{\alpha}$, and of deuterons $X_D$.}
\end{figure}

The phase diagram of nuclear matter, in terms of the composition regarding light and heavy nuclei is depicted in Fig.~\ref{fig_pdyp}. The mass fraction of a particle $i$ is defined by $X_i=A_i n_i/n_B$. In Fig.~\ref{fig_pdyp} we show the contour lines for a mass fraction $10^{-4}$ of the light nuclei $X_a=\sum_{A,Z\leq5}A n\az/n_B$ and of the heavy nuclei $X_A=\sum_{A,Z\geq6}A n\az/n_B$. At temperatures above $1-2$ MeV and low densities nuclear matter consists almost only of free nucleons. Between $10^{-7}$ and $10^{-5}$ fm$^{-3}$ light clusters begin to appear. As will be seen from Fig.~\ref{fig_xlight}, discussed in more detail later, these first light clusters are mainly deuterons. The more symmetric the system is, the earlier is the onset of the light clusters in form of the isospin symmetric deuterons. For all proton fractions some light clusters are present up to $n_B\sim1/2~n_B^0$ where uniform nuclear matter is reached. Only for temperatures below $1-2$ MeV the system consists almost entirely of heavy nuclei. 

At the transition to uniform nuclear matter the following observations can be made: At low temperatures, $T \leq 2$ MeV, the transition density where uniform nuclear matter is reached is increasing with the proton fraction from $n_B=0.3n_B^0$ to $0.7n_B^0$, similar as in the bulk nuclear matter phase diagram of Fig.~\ref{fig_bulkpd}. At the highest temperatures studied, the uniform nuclear matter appears only slightly below saturation density, and the transition density is almost independent of $Y_p$. Furthermore, for $Y_p=0.1$ and $0.01$ we find that the transition shifts to significantly larger densities above a certain temperature at which the heavy nuclei disappear.

The phase diagram of heavy nuclei in Fig.~\ref{fig_pdyp} can be seen as a manifestation of the liquid-gas phase transition of bulk nuclear matter, Fig.~\ref{fig_bulkpd}. The critical temperature up to which heavy nuclei are abundant increases from roughly 2 MeV at $Y_p=0.01$ up to 20 MeV for symmetric nuclear matter. Obviously, the presented phase diagram depends on the somewhat arbitrary distinction between light and heavy nuclei by the proton number $Z\leq 5$. For example for $T \geq 10$ MeV, the heavy nuclei have actually only very low mass and charge numbers. The appearance of these intermediate nuclei leads to the broad extension of $X_A$ at high temperatures in Fig.~\ref{fig_pdyp}. Nevertheless, if one takes the pecularities of the different models into account, there is a qualitative agreement with the phase diagrams of e.g.~Refs.~\cite{lattimer91,shen98_2, muller95}.

\begin{figure}
\includegraphics{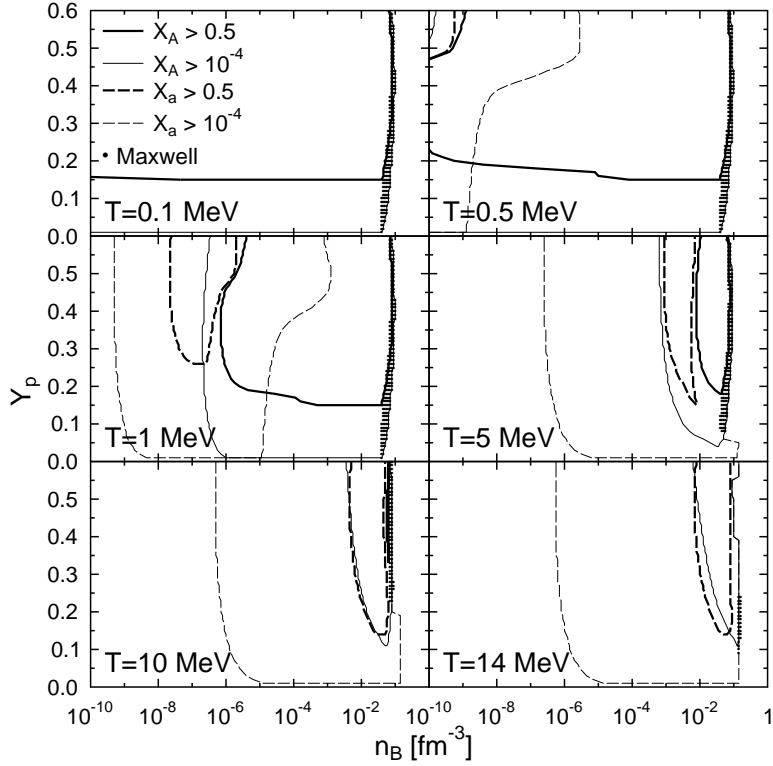}
\caption{\label{fig_pdt}The phase diagram of nuclear matter for different temperatures in the $Y_p-n_B$ plane. Thick (thin) solid lines enclose the region where the mass fraction $X_A$ of heavy clusters with $Z \geq 6$ exceeds 0.5 ($10^{-4}$). The dashed lines correspond to the mass fractions of light clusters $X_a$ with $Z \leq 5$. Dots mark the densities which lie in the mixed phase of the Maxwell-transition from non-uniform to uniform nuclear matter.}
% For every $Y_p$ the point at the highest density marks the beginning of uniform nuclear matter.}
\end{figure}
Fig.~\ref{fig_pdt} shows again the phase diagram, but this time in the $Y_p-n_B$ plane for some selected temperatures. Contour lines for a mass fraction of 0.5 are shown by the thick lines. With this criterion the dominant phase can directly be identified. For all temperatures, nucleons are the most abundant component for proton fractions below $\sim 0.1$. In this case there are only few protons in the system and thus only a small amount of nuclear clusters can form. For larger $Y_p$ the phase diagrams show a strong temperature dependence. At the lowest temperature $T=0.1$ MeV, as expected, the heavy nuclei fill the rest of the $Y_p-n_B$ plane up to $\sim 1/2~n_B^0$ where uniform nuclear matter is reached. At a temperature of 0.5 MeV a small region in the upper left corner appears which is dominated by light clusters. At such low densities the heavy nuclei are dissolved into lighter clusters, and as these light clusters are mainly $\alpha$-particles (see Fig.~\ref{fig_xlight}) because of their relatively strong binding, this happens only at very large proton fractions of $\sim 0.5$. At a temperature of 1 MeV this light cluster region is shifted to higher densities. Again, the light clusters are mainly $\alpha$-particles which explains their favorable appearance around $Y_p\sim 0.5$. At the lowest densities even the light clusters are dissolved into free nucleons. At densities larger than $10^{-6} - 10^{-5}$ fm$^{-3}$ the heavy nuclei dominate until uniform nuclear matter is reached. From $T=5$ to 14 MeV light and heavy nuclei appear only in a very narrow density band between $10^{-3}$ and 0.1 fm$^{-3}$. The region dominated by heavy nuclei shrinks with increasing temperature. For $T=14$ MeV the mass fraction of heavy nuclei never exceeds $0.5$.

In Fig.~\ref{fig_pdt} also the transition to uniform nuclear matter is illustrated. The dots show the density-grid-points of the calculation which are based on the Maxwell-construction, as explained earlier. We find that with increasing temperature the mixed phase region becomes smaller. It even disappears completely for very low $Y_p$ and temperatures $\geq 5$ MeV, because then the mixture of nuclei and nucleons behaves almost like uniform nuclear matter. Due to the same reason the transition is shifted to larger densities. For $T\leq 5$ MeV, where many heavy nuclei exist, the mixed phase becomes most extended at low $Y_p\sim 0.2$. On the contrary, at larger $Y_p$ the Maxwell transition region becomes narrower.

\begin{figure}
\includegraphics{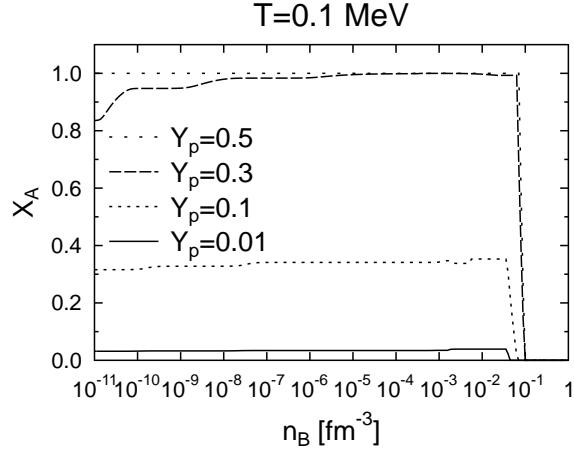}
\caption{\label{fig_xat0.1}The mass fraction of heavy nuclei $X_A$ as a function of the baryon number density $n_B$ for a temperature $T=0.1$ MeV and different proton fractions $Y_p$.}
\end{figure}
\begin{figure}
\includegraphics{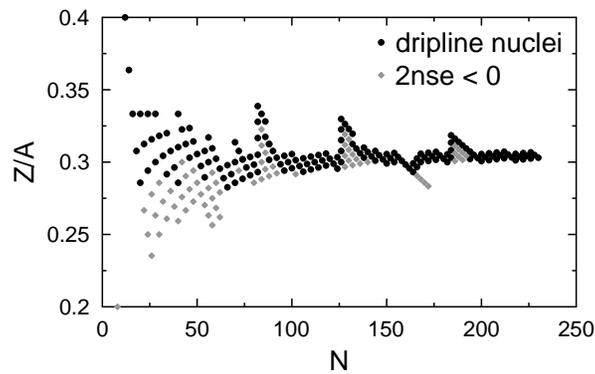}
\caption{\label{fig_dripyp}The charge to mass ratio $Z/A$ as a function of the neutron number $N$ of nuclei which lie on the neutron drip line (black circles) and which have a negative two neutron separation energy, i.e.~which lie behind the neutron drip line (grey diamonds).}
\end{figure}
The independence of the phase diagram on the density at $T=$ 0.1 MeV is further analyzed in Fig.~\ref{fig_xat0.1}. There the mass fraction of heavy nuclei is shown as a function of density for various proton fractions. For such low temperatures the mass fraction of heavy nuclei is almost constant throughout all densities. This can be explained in the following way: For all shown values of $Y_p$ free protons are never present and therefore all the protons are concentrated in nuclei. For proton fractions $Y_p \geq 0.3$ the system consists almost only of heavy nuclei. Even though the neutron chemical potential increases slowly with density, the neutron drip ($\mu_n=m_n$) is never reached and as the temperature is low the free neutron density remains vanishingly small. There will be a critical $Y_p^{drip}$, below which the neutron drip occurs, with $Y_p^{drip} \sim 0.31$ in our calculations. For proton fractions below this critical value a dilute free neutron gas with $\mu_n \simeq 0$ appears, which leads to the drastic reduction of $X_A$. Under the condition $\mu_n \simeq 0$ exclusively such nuclei are being populated, whose two-neutron separation energy are close to zero, which means that they are neutron drip nuclei. Fig.~\ref{fig_dripyp} shows the charge to mass ratio of nuclei which lie on the drip line and of those whose two-neutron separation energy is negative. They all have $Z/A \sim 0.3$. As no free protons are present, the mass fraction of heavy nuclei is directly determined by the total proton fraction, $X_A \sim Y_p/0.3$ for $Y_p<0.31$, and is independent of density, which is in good agreement with the results of Fig.~\ref{fig_xat0.1}.

\begin{figure}
\includegraphics{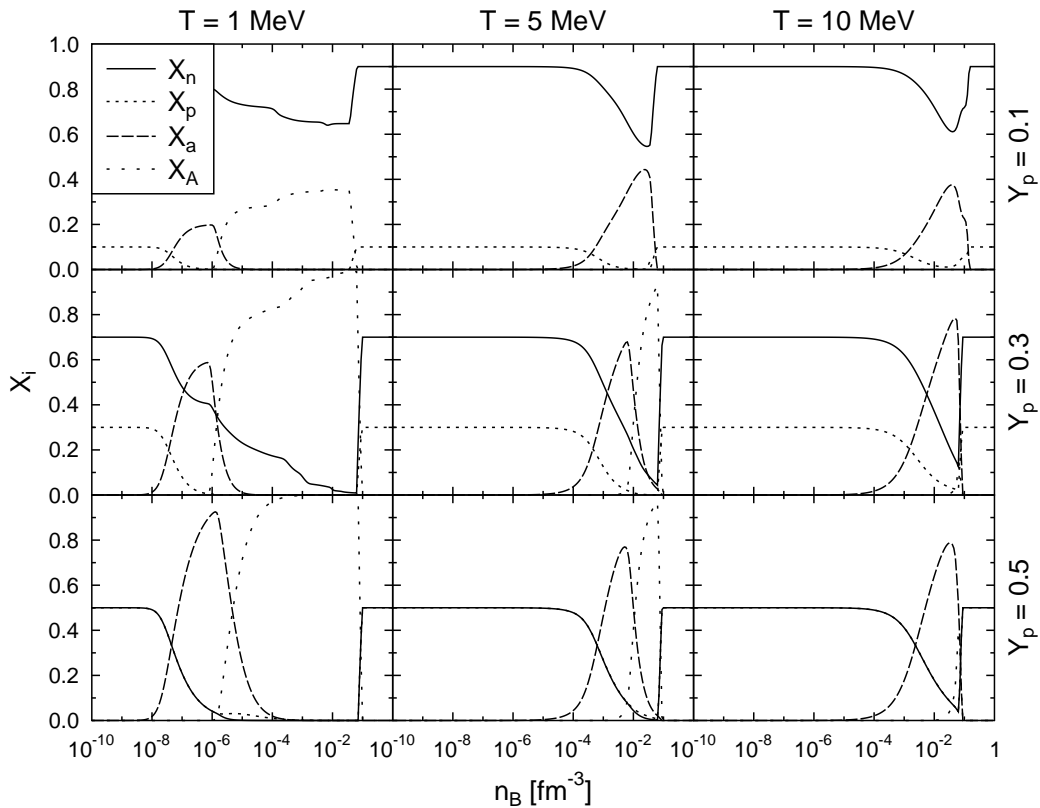}
\caption{\label{fig_xi}The mass fraction of free neutrons $X_n$, free protons $X_p$, light nuclei $X_a$, and heavy nuclei $X_A$, as a function of the baryon number density $n_B$. The columns show the different temperatures $T=1, 5,$ and 10 MeV (from left to right), the rows the proton fractions $Y_p=0.1, 0.3,$ and 0.5 (from top to bottom).}
\end{figure}
For higher temperatures the composition changes significantly. Depending on the actual values of temperature, density and proton fraction, free protons, free neutrons and light and heavy nuclei appear in different concentrations. Fig.~\ref{fig_xi} demonstrates that for temperatures of $T=1$ MeV and densities up to $n_B \sim 10^{-8}$ fm$^{-3}$ mainly only free nucleons are present. At larger densities the protons cluster together to form light nuclei and thus the free proton density vanishes. The light clusters tend to be symmetric and thus the fraction of the free neutrons is reduced by the same value as the one for protons. Due to the same reason, the maximum mass fraction of the light nuclei is roughly twice the proton fraction. At densities larger than $10^{-6}$ fm$^{-3}$ heavy nuclei appear and replace the lighter nuclei. With increasing density the nuclei grow in size and become more asymmetric so that more neutrons are bound in nuclei. The fraction of heavy nuclei increases further and becomes close to 1 for large proton fractions. As there are no nuclei with $Z/A< 0.1$ in the mass table, some free neutrons have to remain for $Y_p=0.1$. The stepwise change of the fractions which can be seen for $Y_p=0.1$ and 0.3 can be attributed to transitions between different nuclei which give the main contribution to the composition.

At a temperature of $T=5$ MeV the free nucleon regime extends up to $n_B \sim 10^{-4}$ fm$^{-3}$. At larger densities the nucleons are successively replaced by light nuclei. For larger proton fractions there are sufficiently many protons in the system that finally all the nucleons can be bound to nuclei. Only in these cases a significant contribution of the heavy nuclei appears, shortly before uniform nuclear matter is reached. At a temperature of 10 MeV the overall composition looks similar. The onset of the light nuclei takes place at roughly the same density, but their presence extends up to higher densities. For $T=10$ MeV heavy nuclei play only a minor role. Only for large $Y_p$ heavy nuclei appear at all, and then only at densities slightly below the transition to uniform nuclear matter.

\begin{figure}
\includegraphics{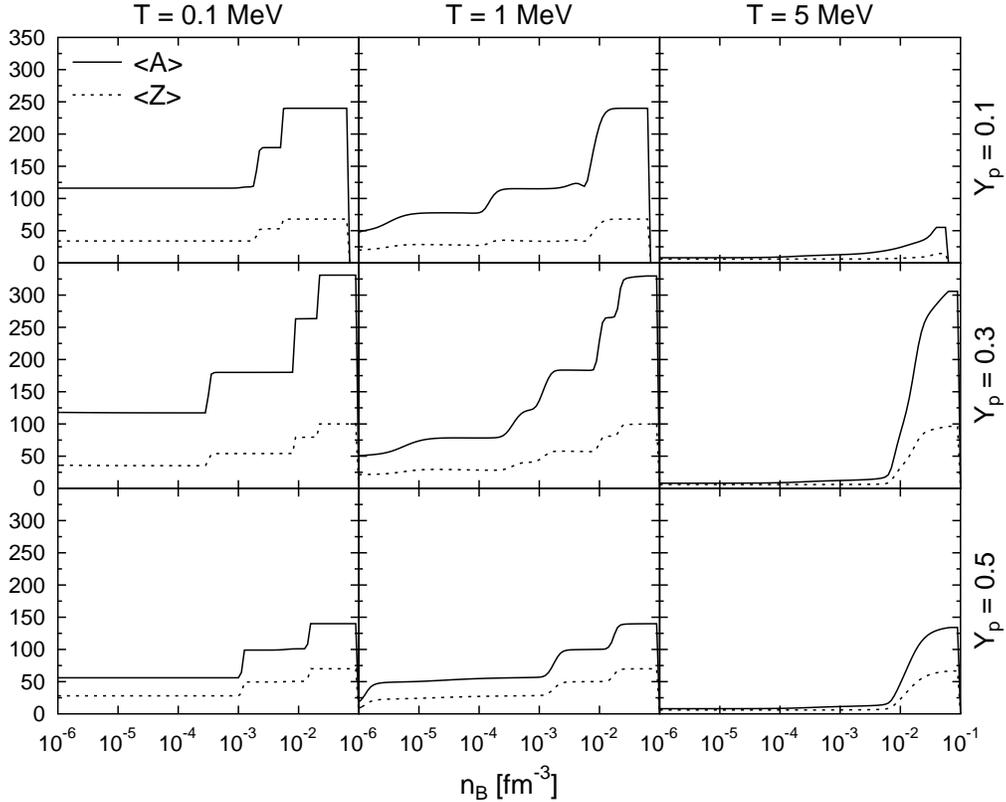}
\caption{\label{fig_za}The average mass and proton numbers $<A>$ and $<Z>$ of heavy nuclei with $Z \geq 6$.}
\end{figure}
\begin{figure}
\includegraphics{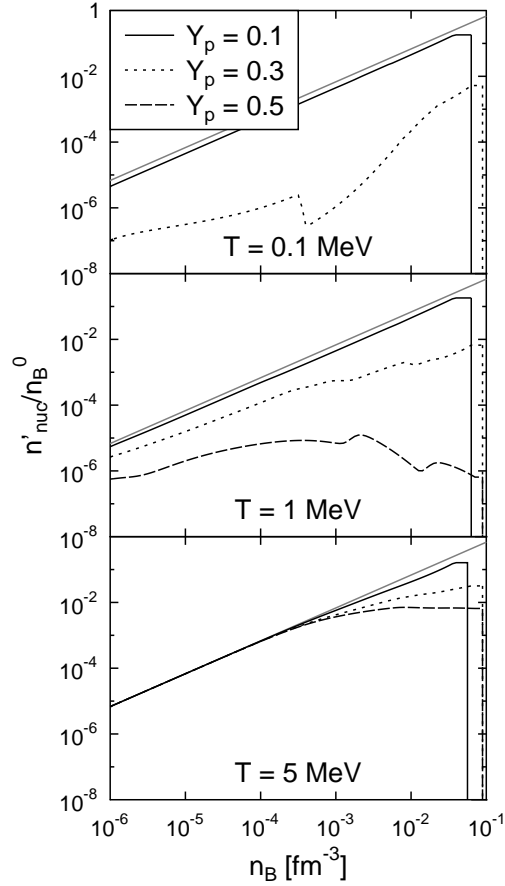}
\caption{\label{fig_freen}The free local nucleon number density $n'_{nuc}=n'_n+n'_p$ outside of nuclei in units of the saturation density $n_B^0=0.147$ fm$^{-3}$. Grey lines show the total baryon number density.}
\end{figure}
The chemical composition regarding the average mass and proton number of heavy nuclei is further analyzed in Fig.~\ref{fig_za}. Temperatures and densities are shown for which their mass fraction is large (see Figs.~\ref{fig_xat0.1} and \ref{fig_xi}). The first thing to note is the stepwise increase of $<A>$ and $<Z>$ for $T=0.1$ and 1 MeV, which was already seen in Fig.~\ref{fig_abeta} before. For such small temperatures the distributions of nuclei are narrow and $<A>$ and $<Z>$ are mainly given by one single nucleus. This causes also the steps in the mass fractions observed in Fig.~\ref{fig_xi} for $T=1$ MeV. Shells effects are strong, as it comes out that most of these nuclei have neutron magic numbers 28, 50, 82, 126 or 184. This is in strong contrast to models which are based on the Thomas-Fermi approximation \cite{shen98, shen98_2} or a liquid-drop parameterization \cite{lattimer91}, which are not able to reproduce any shell effects. In these models the mass and charge number change continuously. By looking at the different values of $Y_p$ shown in Fig.~\ref{fig_za}, we find that the largest nuclei appear for $Y_p=0.3$. For $T=0.1$ and $1$ MeV a similar composition is found, differences appear only at low densities. For $T=5$ MeV and densities below $10^{-2}$ fm$^{-3}$, where almost no heavy nuclei exist, the average heavy nucleus is $^8$C, because it is the lightest nucleus with $Z=6$. At larger densities, when the fraction of heavy nuclei increases, the nuclei grow in size. For this large temperature we observe a continuous change of the mass and charge number, indicating less pronounced shell effects and broad distributions.

In the present work the shell structure of nuclei is not modified by the medium. The results of Ref.~\cite{buervenich07} show that the impact of the dense electron gas on nuclear properties is rather small. To estimate the role of free nucleons outside of the nuclei, their local number density $n'_{nuc}=n'_p+n'_n$ is depicted in Fig.~\ref{fig_freen}. Only for $Y_p = 0.1$ or at a temperature of 5 MeV the free nucleon density exceeds $0.01 n_B^0$. In the latter case heavy nuclei are only abundant between $10^{-2}$ fm$^{-3} < n_B<10^{-1}$ fm$^{-3}$. At larger temperatures the free nucleon density increases further, but then the heavy nuclei only play a minor role, see Fig.~\ref{fig_xi}. At lower temperatures, more heavy nuclei exist in a broader range of density. But then the nucleon density is only significantly large, if the proton fraction is very low. In typical supernova simulations the proton fraction for most of the matter is actually rather high, $0.3<Y_p$, supporting the neglect of the medium modifications of the nuclei due to the unbound nucleons.

\begin{figure}
\includegraphics{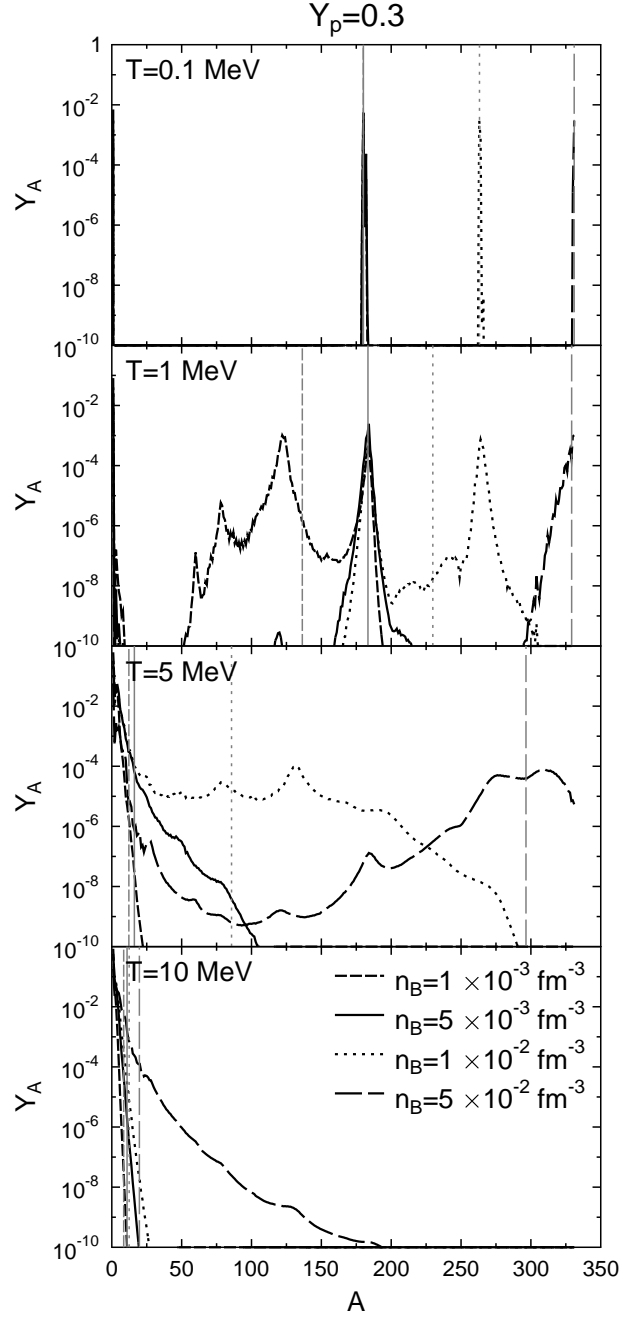}
\caption{\label{fig_ya}Mass distributions for selected temperatures and densities for $Y_p=0.3$. Vertical lines show the average mass number $<A>$ of the heavy nuclei with $Z\geq6$.}
\end{figure}
\begin{figure}
\includegraphics{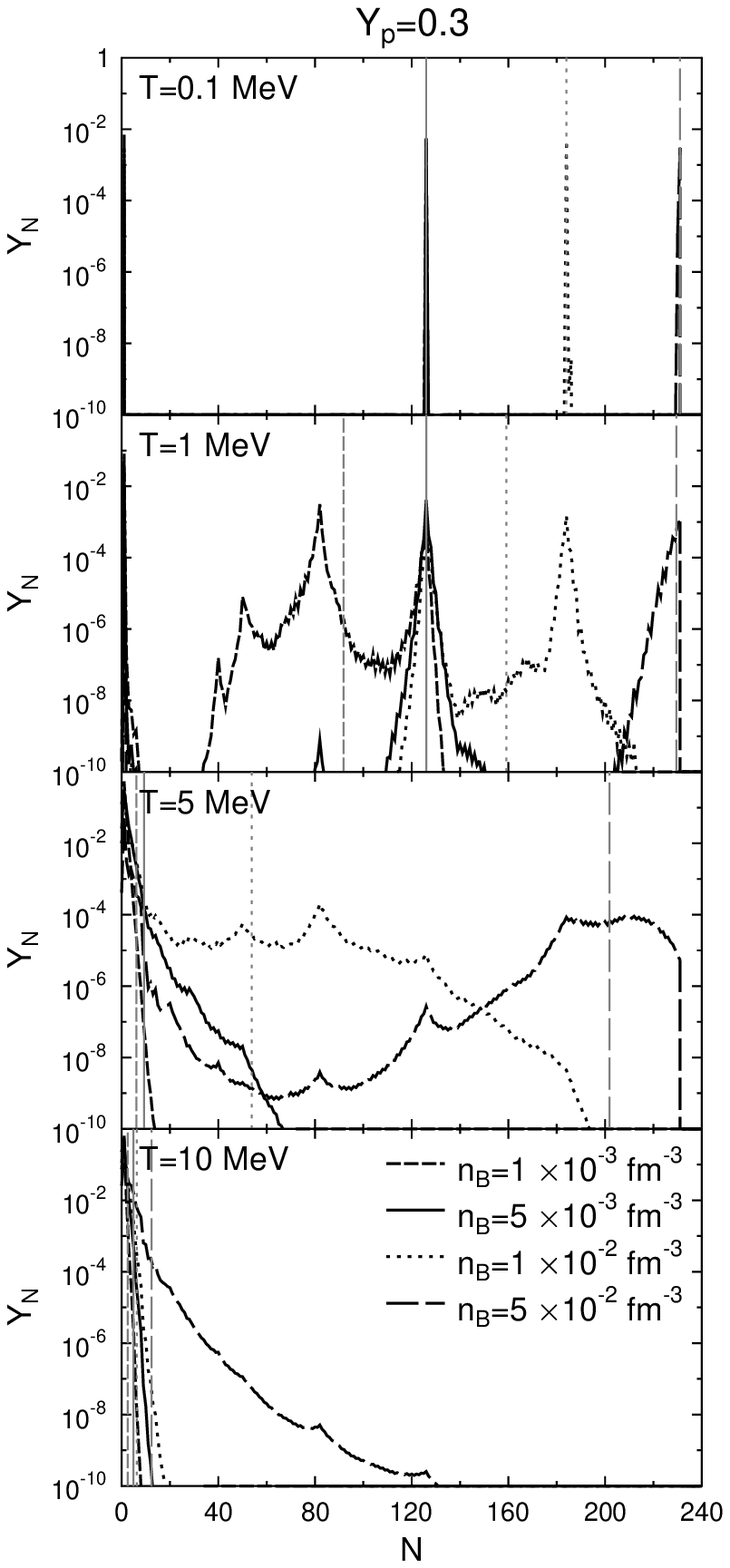}
\caption{\label{fig_yn}As Fig.~\ref{fig_ya}, but now for the distribution of the neutron number $N$.}
\end{figure}
Fig.~\ref{fig_ya} depicts the distributions of nuclei with respect to the mass number $A$. Here we are showing relative yields $Y_A=\sum_Z n\az / \sum\az n\az$. At a temperature of 0.1 MeV the distributions are sharply peaked. We note that the distributions at $n_B=10^{-3}$ fm$^{-3}$ and $5\times 10^{-3}$ fm$^{-3}$ lie on top of each other for this temperature. The mean value $<A>$ of the heavy nuclei with $Z\geq6$ coincides with the peak of the distribution. In this case the single nucleus approximation should give almost identical results compared to the NSE description. In Fig.~\ref{fig_yn}, which shows the neutron number distribution, one can see that the dominant nuclei at $T=0.1$ MeV have the neutron magic numbers 126 or 184. At $n_B=5\times 10^{-2}$ fm$^{-3}$ the dominant nucleus is already at the border of the nuclear mass table and therefore no neutron magic number can be identified. At $T=1$ MeV temperature effects become visible and the distributions broaden. The magic nuclei mentioned before ($N=126$ and 184) are still strongly populated, but additional peaks appear. E.g.~for $n_B=10^{-3}$ fm$^{-3}$ the strong peaks can be identified with the neutron magic numbers 50, 82 and 126. Nuclei with $N=40$ also seem to be rather strongly bound in the model TMA. In general most of the peaks in the mass distributions can be assigned to neutron magic numbers. As was already found in Ref.~\cite{ruester06} for the outer crust of neutron stars, proton magic numbers do not play a significant role. The proton number determines the Coulomb energy of the nuclei. Thus the charge of the nuclei can not be adjusted as freely as their neutron number. Although for $T=1$ MeV the distributions are still sharply peaked, because several peaks with similar yields appear, it would not be appropriate to use the mean values to describe the charge and mass distributions. E.g.~at $n_B=10^{-2}$ fm$^{-3}$ the distribution shows two similar maxima, with the mean value $<A>$ lying in between. Compared to statistical models which are based on a liquid-drop formulation without shell corrections the typical Gaussian distributions are not found in the present work because the distributions are dominated by shell effects. In Ref.~\cite{botvina08} a shell correction was included which resulted in a similar delta-function like distribution. At a temperature of 5 MeV, which is larger than the typical energy associated with shell effects, the neutron magic numbers are still visible, but much weaker. At large densities the distributions become very broad and extend over the whole nuclear chart. With increasing density the typical behavior expected at the liquid-gas phase transition line, compare with Fig.~\ref{fig_bulkpd}, can be identified, as it was also discussed in Ref.~\cite{botvina08}. For $n_B=10^{-3}$ fm$^{-3}$ the distribution is an exponential. With increasing densities it changes to a flattening power-law. Finally at $n_B=5 \times 10^{-2}$ fm$^{-3}$ the distribution has the typical U-shaped form. For $T=10$ MeV mainly light clusters are populated and the distributions are exponential. For $T=10$ MeV only for the largest density the distribution is power-law like, again an indication for the onset of the liquid-gas phase transition. At this large temperature shell effects are almost not visible any more.

\begin{figure}
\includegraphics{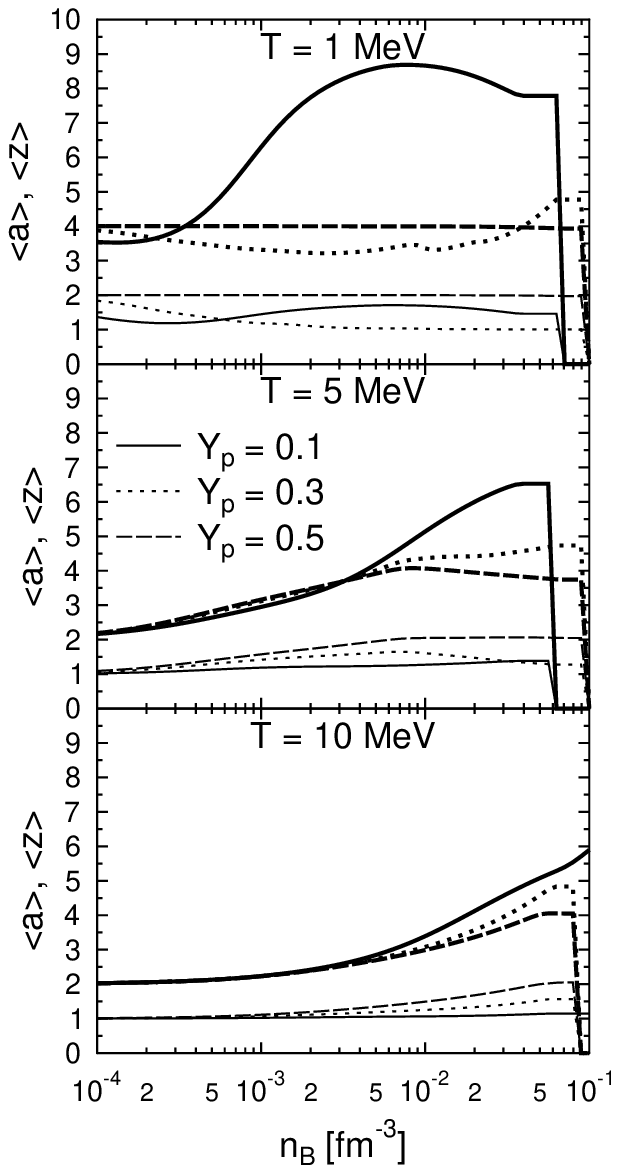}
\caption{\label{fig_zalight}The average mass number $<a>$ (thick lines) and average proton number $<z>$ (thin lines) of the light nuclei with $Z\leq 5$.}
\end{figure}
The contribution of the light clusters is further analyzed in Fig.~\ref{fig_xlight}. At a temperature $T=1$ MeV the light clusters are mainly $\alpha$-particles in the region where they appear in large fractions. With increasing temperature lighter particles are favored, leading to an increase of the deuteron fraction and a reduction of the $\alpha$-particles. For $T=10$ MeV the deuteron mass-fraction surmounts the $\alpha$ mass-fraction at almost all densities. For $T=5$ and 10 MeV and densities below $10^{-4}$ fm$^{-3}$ the light clusters are almost exclusively deuterons, but at these densities the light cluster fraction is relatively small, $X_a<0.01$. For the same temperatures but larger densities from $n_B=10^{-4}$ fm$^{-3}$ to saturation density not only alphas and deuterons are important, but rather the whole distribution of light nuclei. The average mass and charge number of the light clusters are depicted in Fig.~\ref{fig_zalight}. Note that for $T=1$ MeV the light nuclei fraction is actually small for the density range which is shown in Fig.~\ref{fig_zalight}. For symmetric matter at $T=1$ MeV, the average light cluster is well represented by $^4$He. For $Y_p=0.3$, above $10^{-4}$ fm$^{-3}$ the average mass $<a>$ and charge $<z>$ are in general smaller. Close to the transition to uniform nuclear matter very neutron-rich hydrogen isotopes are formed. The contribution of light, very asymmetric nuclei which form inside the free neutron gas for $Y_p=0.1$ leads to the second increase of the light cluster fraction seen in Fig.~\ref{fig_xlight} for $T=1$ MeV and $Y_p=0.1$. For $T=5$ the light clusters are mainly deuterons at low densities. At larger densities, $<a>$ and $<z>$ again behave differently for different $Y_p$. For $Y_p=0.1$ predominantly light clusters with low charge $Z=1$ appear. With increasing density these hydrogen-isotopes become heavier and more asymmetric, with $<z>/<a> \sim 0.2$ before matter becomes uniform nuclear matter. With increasing $Y_p$ clusters with a higher charge are populated which are more isospin-symmetric. Thus with increasing density, the mass does not increase as much as for low $Y_p$. The average mass and charge number for $T=10$ MeV look similar as for 5 MeV, but the distributions are shifted to higher densities. Only above $10^{-3}$ fm$^{-3}$ the deuterons are replaced by heavier particles. Again, the clusters become more symmetric and have larger proton numbers but lower mass when the proton fraction increases.

\subsection{Equation of State}
\begin{figure}
\includegraphics{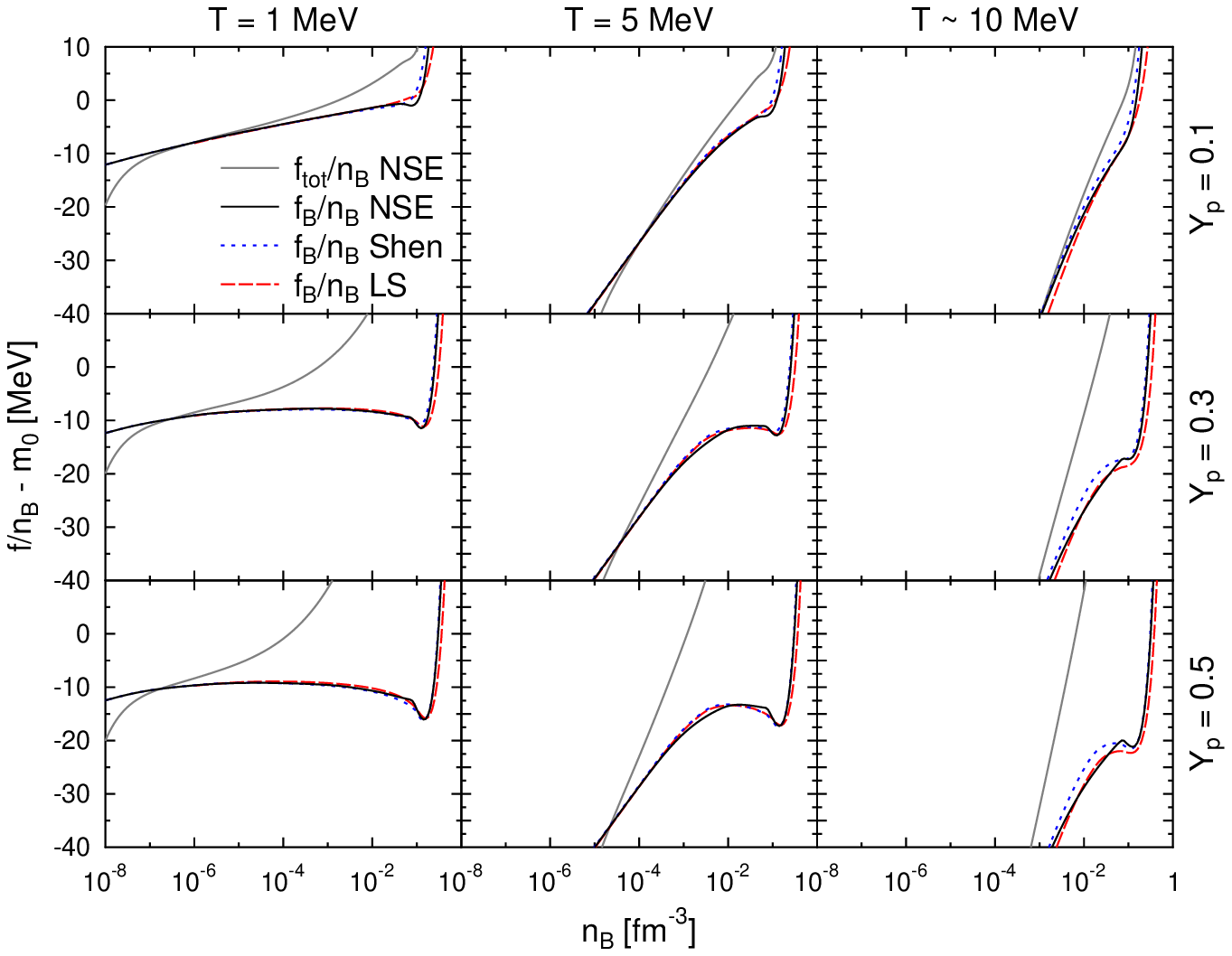}
\caption{\label{fig_f}The baryonic contribution to the free energy (solid black lines) and the total free energy (solid grey lines) per baryon with respect to the rest mass. The model used in the present investigation (NSE) is compared to the results of Shen et al.~\cite{shen98,shen98_2} (blue dotted lines) and of Lattimer and Swesty (LS) \cite{lattimer91} (red dashed lines). (color version online)}
\end{figure}
The thermodynamic potential for given $(T,n_B,Y_p)$ is the Helmholtz free energy and all other thermodynamic quantities are derived from it. In Fig.~\ref{fig_f} the total free energy density (including baryons, electrons/positrons and photons) is depicted. As the electron, positron and photon contribution is trivial, we also show the baryonic part of the free energy. We compare it to the results of LS \cite{lattimer91} and Shen et al.~\cite{shen98, shen98_2}. We do not use the routine of the LS EOS but their table for the potential model SkM* \cite{bartel82}, which is publicly available online. For the LS EOS the temperature of $10.67$ MeV is shown, because no entry for $T=10$ MeV exists in the chosen table and we do not want to use any interpolation here. We note that the Shen EOS has a higher incompressibility, $K=$281 MeV, and symmetry energy, $a_{sym}=36.9$ MeV, than the LS EOS, which has $K=$217 MeV and $a_{sym}=31.4$ MeV. Thus the Shen EOS represents a stiffer EOS with a higher maximum mass for a cold deleptonized neutron star, $M_{max}=2.2$ M$_\odot$ \cite{shen98}, than the LS EOS, $M_{max}=1.62$ M$_\odot$ \cite{stone07}. 

It is important that the three different models are based on very different model assumption for the description of non-uniform nuclear matter, as described in the introduction. Furthermore, they use different models for the nuclear interactions with different nuclear matter properties (e.g.~saturation density, compressibility, symmetry energy). For the shown temperatures and proton fractions, up to densities of $\sim 10^{-4}$  fm$^{-3}$ the free energies of the three models are almost identical. Above saturation density the different properties of uniform nuclear matter become visible. The RMF model TMA used in the statistical model is more similar to the Shen EOS, which is also based on a RMF model, but on the different parameterization TM1. In the intermediate density range the differences in Fig.~\ref{fig_f} are small and of similar size as the differences between the EOSs of LS and Shen. It is a surprising result that the present, `non-microscopic' model is able to give a reasonable description regarding the equation of state across all densities.

One certain feature of the NSE description can be observed at large temperatures, e.g.~at $T=5$ MeV: Although the free energy of uniform nuclear matter is rather large, the free energy is lower than in both of the two single nucleus approximation-models at $10^{-3}$  fm$^{-3}<n_B<10^{-2}$  fm$^{-3}$. As can be seen in Fig.~\ref{fig_ya} the distributions develop from a steep exponential to a very flat power-law shape in this density region. At the beginning of this transition the light clusters become very abundant, see Fig.~\ref{fig_xlight}. Besides a large fraction of $\alpha$-particles and deuterons all of the light clusters contribute to the composition. Later we will give further evidence that it is the contribution of light clusters in the NSE model which leads to the reduction of the free energy as seen in Fig.~\ref{fig_f}. In the other two EOSs only $\alpha$-particles are considered and this behavior can not be observed. However, as was shown in Fig.~\ref{fig_freen}, the free nucleon density is rather large under these conditions, so that medium effects could lead to changes in the composition. We will address this aspect further in the conclusions of section \ref{sec_sum}. For a low temperature of 1 MeV there are no systematic differences between the three different models. As can be seen from Fig.~\ref{fig_xlight} here the light clusters are very well described by $\alpha$-particles, which are included in all three models. At $T=10$ MeV the differences of the different models are in general more pronounced. The lowered free energy of the statistical model is still present, but shifted to slightly larger densities $\sim 10^{-2}$ fm$^{-3}$. 

At densities larger than $\sim 10^{-2}$  fm$^{-3}$ the statistical model has a higher free energy than the other two models. Here the nuclear mass table and the description of the transition to uniform nuclear matter is too restrictive, as the nuclei can not grow arbitrary in size and are limited in $Z/A$. The kinks in the baryonic contribution which are visible around $\sim 10^{-1}$  fm$^{-3}$ come from the Maxwell-construction which is used here. A Gibbs-construction in which the requirement of local proton fraction conservation would be replaced by global conservation of the proton fraction (as discussed in Sec.~\ref{ss_trans}) would lead to a more continuous transition with an earlier onset of the uniform nuclear matter phase and a lower free energy. At large temperatures $T\geq10$ MeV and very low $Y_p$ the transition is smoother, as expected from the discussion of Fig.~\ref{fig_pdt}, because in this case the contribution of nuclei is low and the non-uniform matter phase behaves very similar to uniform nuclear matter. In contrast, for $Y_p=0.5$ and $T=1$ MeV the kink in the baryonic free energy becomes most strongly pronounced. However, these kinks disappear in the total free energy per baryon, when the electrons are added to the baryons. The total free energy and its first derivatives behave continuously, as discussed before and as we will also show in the following.

\begin{figure}
\includegraphics{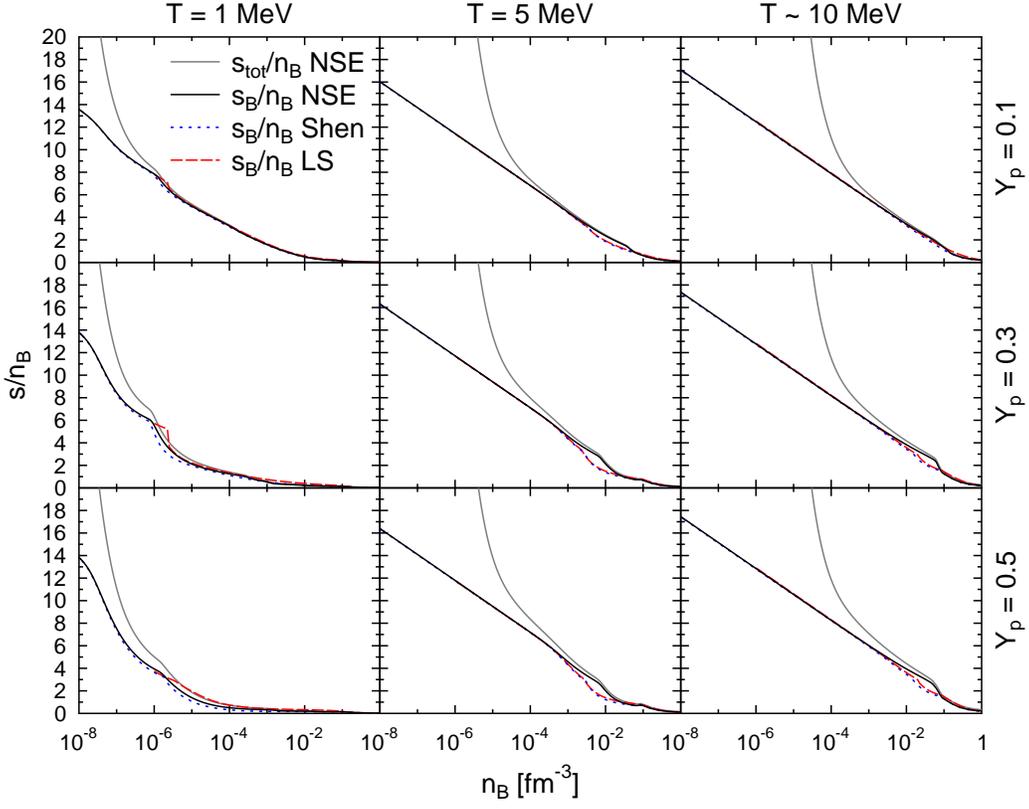}
\caption{\label{fig_s}As Fig.~\ref{fig_f}, but now for the entropy per baryon.}
\end{figure}
Fig.~\ref{fig_s} depicts the entropy per baryon. All models give very similar results for $T=1$ MeV. For this low temperature there seems to be a better agreement of the statistical model with the Shen EOS. The bump around $n_B=10^{-6}$ fm$^{-3}$ which is most dominant at $Y_p=0.3$ arises when the large light cluster contribution (mostly alphas) is replaced by heavy nuclei. There the LS EOS shows a rather abrupt change in the entropy. At larger temperatures, the entropy behaves almost like the one of an ideal gas with $s/n_B \propto -ln(n_B)+const.$. Only above $n_B\sim10^{-4}$ fm$^{-3}$ deviations from the ideal gas behavior appear, when light clusters are formed.

For higher temperatures at densities around $\sim 5 \times 10^{-3}$ fm$^{-3}$ for $T=5$ MeV and $\sim 5 \times 10^{-2}$ fm$^{-3}$ for $T=10$ MeV the entropy is significantly higher in the statistical model. As noted before, the whole distribution of light and intermediate clusters is important here and leads to the increased entropy. This increased number of available states is the reason for the lower free energy discussed before.

For comparison also the total entropy is shown in Fig.~\ref{fig_s}. No discontinuities are observed, as expected. The total entropy enables to identify the regions where the nontrivial baryonic contribution is important at all. It is shown only for the statistical model, because the leptons and photons are treated identical in all three models. At densities below $10^{-7}$ fm$^{-3}$ for $T=1$ MeV, $10^{-5}$ fm$^{-3}$ for $T=5$ MeV, and $10^{-4}$ fm$^{-3}$ for $T=10$ MeV, the electron-positron plasma determines the entropy almost completely. But at larger densities it is the baryon contribution which gives the largest contribution to the entropy, and electrons, positrons and photons are negligible. In this density range the different descriptions of the different models become important. Thus we conclude that regarding the entropy the different results for the baryonic EOS also affect the total EOS.

\begin{figure}
\includegraphics{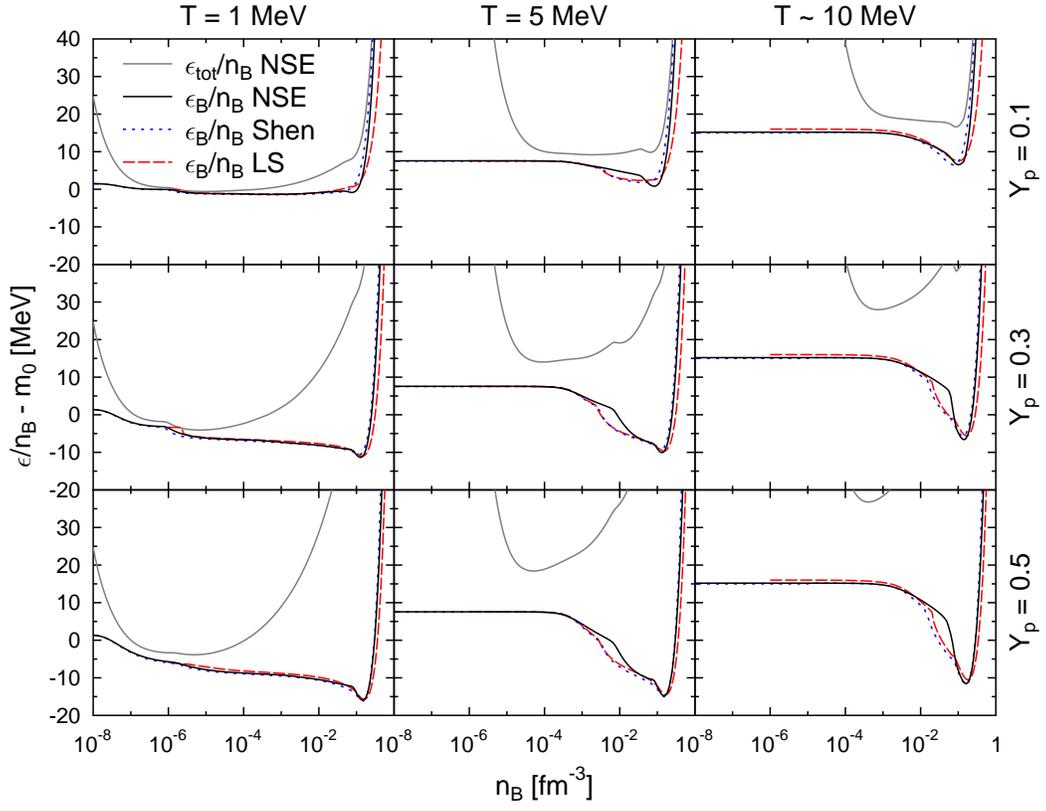}
\caption{\label{fig_e}As Fig.~\ref{fig_f}, but now for the binding energy per baryon.}
\end{figure}
Fig.~\ref{fig_e} shows the binding energy per baryon, which is directly given by the entropy and the free energy ($\epsilon=f+Ts$). In the third column of Fig.~\ref{fig_e} the energy density of the LS EOS is higher at low densities, because it is shown for the slightly larger temperature of $10.67$ MeV, as mentioned before. At lowest densities the ideal gas limit $\epsilon=3/2 n_B T$ is reached in all three models. At $T=1$ MeV the fraction of heavy nuclei becomes already important above $n_B \sim 10^{-8}$ fm$^{-3}$. Their binding energy leads to a decrease of the baryonic energy density. At $T=5$ and $T=10$ MeV the nuclear interactions become visible above $n_B\sim10^{-4}$  fm$^{-3}$. In general, the maximum binding energy is achieved close to saturation density.

In the NSE model, around $10^{-2}$  fm$^{-3}$ for $T=5$ MeV and $5 \times 10^{-2}$  fm$^{-3}$ for $T=10$ MeV, the slightly lower free energy is more than compensated by the increased entropy and therefore the energy density becomes larger than in the other two EOSs at these densities. The apparent differences are even more significant than for the free energy and the entropy, because in the expression for the energy density, $\epsilon=f+Ts$, the difference in the entropy is multiplied by the temperature. Thus the largest difference in the binding energy is observed for $Y_p=0.5$ and $T=10$ MeV. 

As can be seen from Fig.~\ref{fig_e} the baryons are most important for the total energy density at intermediate and very large densities. Again, at low densities, because of the high temperatures, the electron-positron-plasma gives the largest contribution. At larger densities where the positrons have vanished the electrons become degenerate and their energy density rises faster than the one of the attractive nuclear interactions. Obviously, the number density of electrons and their energy density directly depends on the proton fraction $Y_p$. Thus the baryonic contribution becomes more significant for low $Y_p$. Above saturation density the nuclear interactions become strongly repulsive and take over to dominate the energy density. The bumps in the nuclear binding energy which appear below saturation density can still be identified in the total energy density.

\begin{figure}
\includegraphics{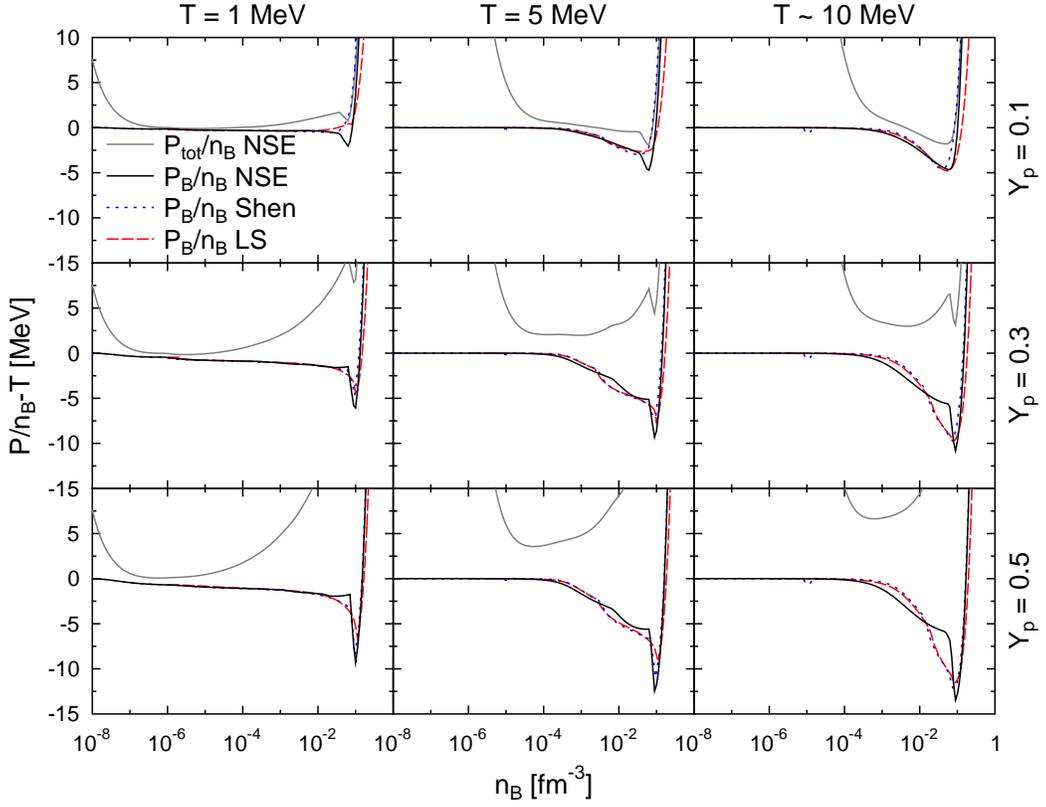}
\caption{\label{fig_pnb}As Fig.~\ref{fig_f}, but now for the pressure divided by the baryon density and with the temperature subtracted, to show the deviations from the ideal gas pressure more clearly.}
\end{figure}
In Fig.~\ref{fig_pnb} the baryonic contribution to the pressure is depicted. It is divided by the baryon density and the temperature is subtracted to see the deviations from the ideal gas pressure more clearly. Presented in this way, the differences of the three different models become very pronounced. The onset of the nuclear interactions appears similarly as in the case for the energy density. At $T=1$ MeV and high proton fractions an increasing fraction of nucleons is bound to heavy nuclei, which grow in size with density, leading to a decreasing pressure. For larger temperatures nuclei become important only at larger densities. The baryonic contribution to the total pressure is important in the same density range as discussed for the energy density. Though the baryonic pressure is negative in many cases, the total pressure is always positive.

At $T=1$ MeV and around $n_B=10^{-1}$ fm$^{-3}$ the pressure increases. Here, matter consists almost exclusively of heavy nuclei. Now the densities are so large that the excluded volume effects significantly increase the pressure of the nuclei, see eq.~(\ref{eq_p}). At even higher densities the ``repulsive'' excluded volume corrections become so strong, that the transition to uniform nuclear matter takes place. The drop in the pressure (most clearly seen for $Y_p=0.5$) arises from the Maxwell construction. In the two other EOSs the transition to uniform nuclear matter is described very differently, and thus the behavior of the baryonic pressure is different, too.

\begin{figure}
\includegraphics{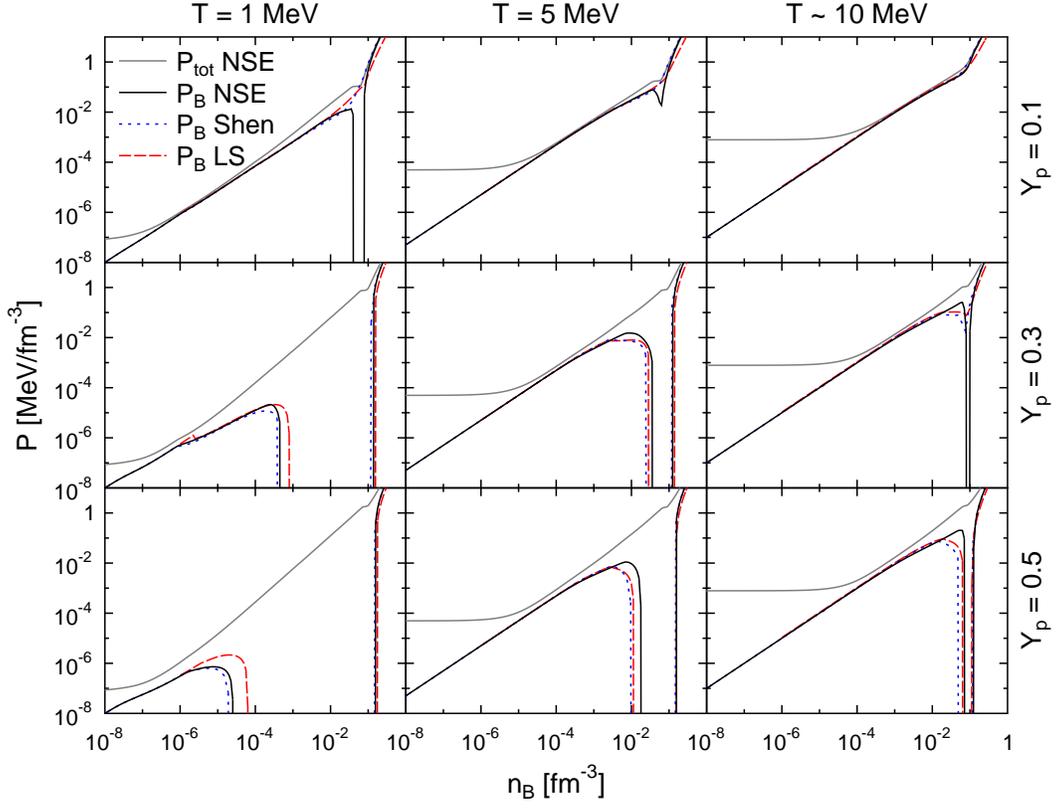}
\caption{\label{fig_p}As Fig.~\ref{fig_f}, but now for the pressure.}
\end{figure}
The baryonic part of the total pressure is depicted in Fig.~\ref{fig_p}. Here, the pressure is not divided by the baryon density, as done in Fig.~\ref{fig_pnb}. Now one sees that the total pressure remains constant during the Maxwell transition. This appears as a sharp pressure drop across the transition if not the pressure but $P/n_B-T$ is plotted. The use of a Gibbs-construction with non-locally fixed $Y_p$ would result in a strictly increasing pressure.

\begin{figure}
\includegraphics{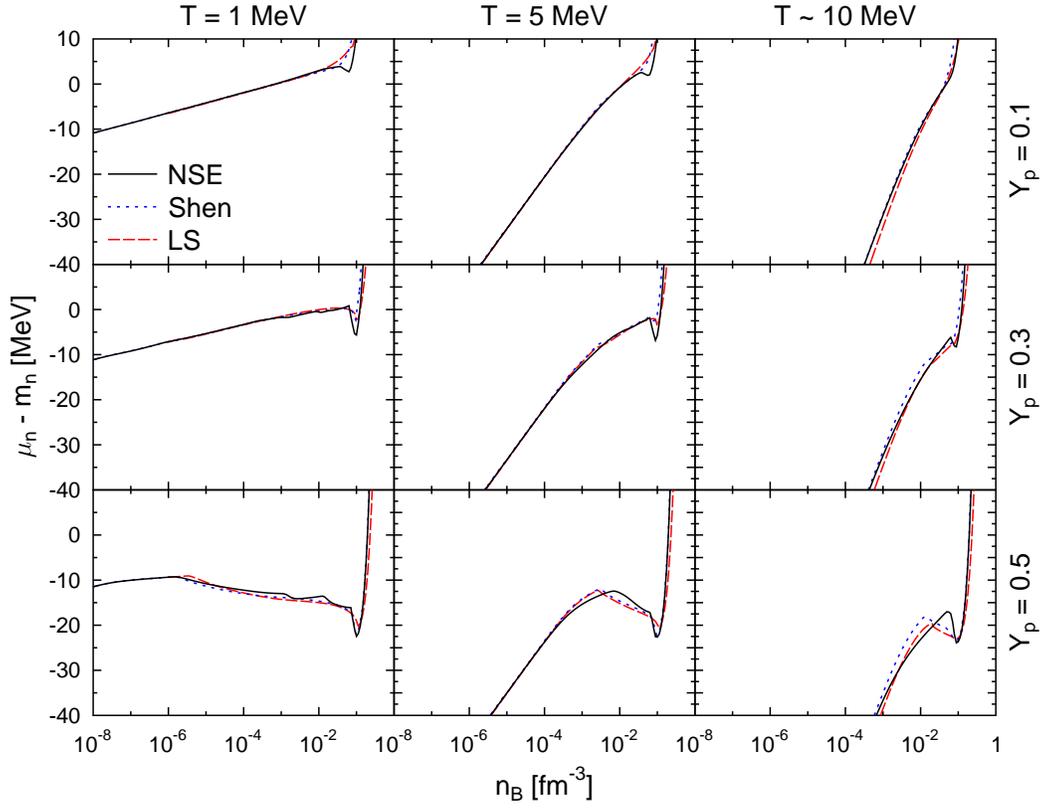}
\caption{\label{fig_mun}As Fig.~\ref{fig_f}, but now for the neutron chemical potential with respect to the neutron rest mass.}
\end{figure}
\begin{figure}
\includegraphics{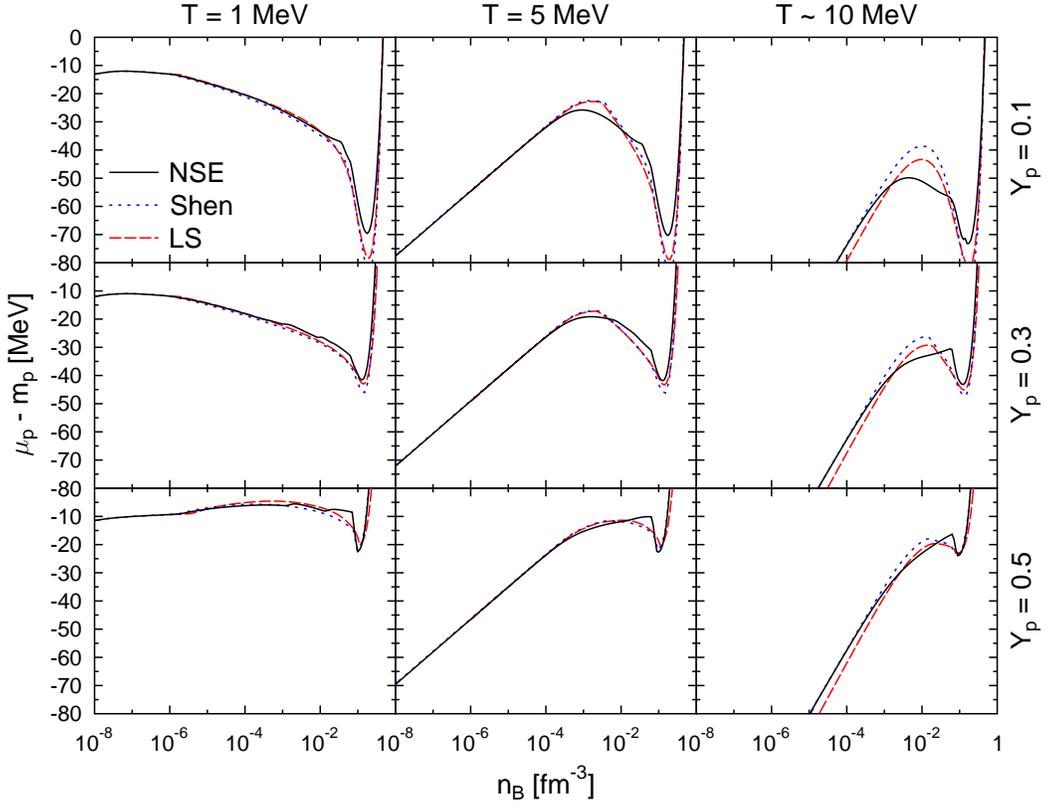}
\caption{\label{fig_mup}As Fig.~\ref{fig_f}, but now for the proton chemical potential with respect to the proton rest mass.}
\end{figure}
The neutron chemical potential is shown in Fig.~\ref{fig_mun}, the proton chemical potential in Fig.~\ref{fig_mup}. At $T=1$ MeV the non-monotonic behavior of the chemical potentials of the NSE model is striking. It comes from the rather discontinuous change in the mass and charge number of the heavy nuclei, as temperature effects are weak, see also Figs.~\ref{fig_za} and \ref{fig_ya}. Besides this, similar results are found as for the other thermodynamic variables. For $T=5$ MeV around $n_B\sim 10^{-3}$ fm$^{-3}$ the chemical potentials are lower, especially the proton chemical potential at low $Y_p$. We attribute this to the strong contribution of the light clusters besides alphas. At $T=10$ MeV this effect happens at $n_B\sim 10^{-2}$ fm$^{-3}$ and is even stronger pronounced. At larger densities the excluded volume effects become important and lead to increased chemical potentials until the phase transition sets in. Here the Maxwell construction is visible as a rather sharp drop in the chemical potentials, especially pronounced for the proton chemical potential. In our mixed phase construction, only the total baryon chemical potential for locally fixed $Y_p$ and local electric charge neutrality, $\mu_B=(1-Y_p)\mu_n+Y_p(\mu_p+\mu_e)$, is equal in the two phases and remains constant across the transition, see Ref.~\cite{hempel09}. The drop of $\mu_n$ and $\mu_p$ across the transition is compensated by the quickly increasing electron chemical potential $\mu_e$.

\begin{figure}
\includegraphics{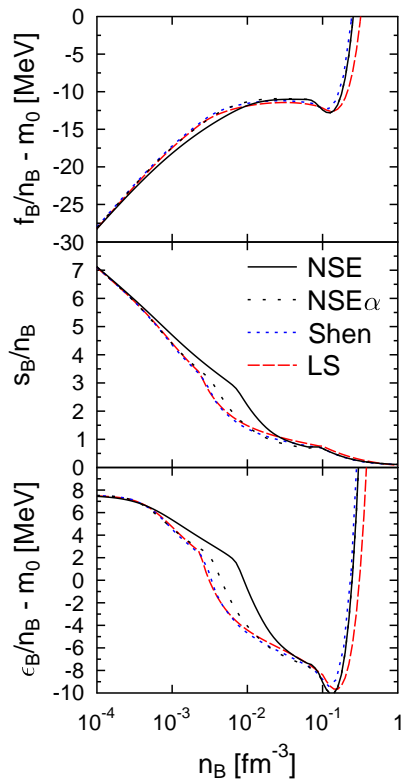}
\caption{\label{fig_nl}The baryonic contribution to the free energy, entropy and energy per baryon as a function of $n_B$ for $T=5$ MeV and $Y_p=0.3$. ``NSE$\alpha$'' shows the results if all light clusters with $A<20$ besides $\alpha$-particles are taken out from the NSE calculation. (color version online)}
\end{figure}
To address the origin of the found deviations of the NSE model from the LS and Shen EOS further, the equation of state is shown if all nuclei with $A<20$ besides nucleons and alphas are taken out in Fig.~\ref{fig_nl}. Now one sees that most of the additional entropy and energy density can indeed be attributed to the light clusters. The decrease of the free energy compared to the other two EOSs between $n_B=10^{-3}$ fm$^{-3}$ and $n_B=10^{-2}$ fm$^{-3}$ is not visible any more. The increase of the free energy before the transition to uniform nuclear matter still remains, as it is caused by different reasons (limitation of $A$ and $Z/A$ because of the use of a mass table). Anyhow, by looking at the energy density and the entropy in Fig.~\ref{fig_nl} one observes that some smaller deviations around $n_B \sim 3\times 10^{-3}$ fm$^{-3}$ remain. The remaining differences are less pronounced and extent over a much smaller range in density. 

\section{Summary \& Discussion}
\label{sec_sum}
In this article we presented a new statistical model for a complete supernova EOS, i.e.~a model which is in principle suitable to describe matter under all typical supernova conditions. In our EOS the interactions of the unbound nucleons are described with a relativistic mean-field model and the parameter set TMA. All nuclei are treated as separate particle species, using the experimental mass table of Ref.~\cite{AudiWapstra} and the theoretical nuclear structure calculations of Ref.~\cite{geng05}, which are based on the same Lagrangian density and parameter set TMA. Excited states of the nuclei are taken into account by a rather simple semi-empirical degeneracy function. Like in most other NSE models, the Coulomb energies are included by the Wigner-Seitz approximation, but are neglected for the protons for simplicity here.  

The different components are set together to a thermodynamic consistent model, based on simple phenomenological and geometrical considerations: We attribute the volume of a hard uniform sphere at saturation density to the nuclei. Thus the nucleons can only reside outside of the nuclei. The nuclei feel also the presence of the unbound nucleons, so that the volume available for them is reduced by all the baryons (inside nuclei and in form of unbound nucleons) in the system. By these assumptions it is assured that nuclei can not exist above saturation density and that the unmodified RMF description is achieved if nuclei are not present. These excluded volume effects are implemented in a thermodynamic consistent way, affecting all thermodynamic quantities. Furthermore, within the model the RMF interactions of the nucleons are coupled to the nuclei via chemical equilibrium.

The thermodynamic consistent description with excluded volume corrections allows to apply the model at all densities and thus also the transition to uniform nuclear matter can be described. Still, the description of nuclei as quasi-particles requires the use of a mixed phase construction, and here we choose the most simple way of a Maxwell construction, based on locally fixed $Y_p$ and local charge neutrality. The phase transition to uniform nuclear matter occurs in most cases at the expected densities between $1/3 n_B^0$ and $2/3 n_B^0$. 

Regarding the composition the following results are found. At high temperatures and low densities the clusters are completely dissolved into an ideal gas of neutrons and protons. With increasing density light clusters form which finally are replaced by heavy nuclei until uniform matter is reached. The proton fraction $Y_p$ sets the maximum fraction of nuclei. We note that in our calculations the chemical potentials of the nuclei were always low enough to describe them by a Maxwell-Boltzmann distribution. Before possible condensates of light clusters could form the light nuclei are replaced by heavy nuclei. Thus we think that for addressing the question of Bose-Einstein condensation in supernova matter it is necessary to include heavy nuclei, too. 

At $T=0.1$ MeV up to the neutron drip only heavy nuclei are present, above which a dilute free neutron gas forms. We compared our composition at $T=0.1$ MeV to the one of Ref.~\cite{ruester06}, and found a very good agreement. Thus we can conclude that a very accurate description in the low temperature and low density limit is achieved. This makes our model particular useful for the description of the evolution of the outer crust of a hot proto-neutron star to a cold deleptonized neutron star. 

The nuclear distributions look qualitatively similar to the ones of Ref.~\cite{ishizuka03}, where a different NSE model is used. At low $T$, due to neutron shell effects, the mass distributions are not the typical Gaussians as in Ref.~\cite{botvina08}, but have rather a delta-function-like shape. Nuclei with the neutron magic number $N=$28, 50, 82, 126 or 184 appear with very high abundance. The narrow distributions lead to a stepwise change of the average mass number $<A>$ and $<Z>$ with density, in contrast to the continuous change in the models of LS \cite{lattimer91} and Shen et al.~\cite{shen98, shen98_2}. Consequently, also the mass fractions of the different components show a rather discontinuous behavior. Though the sharply peaked distributions support the single nucleus approximation (SNA), we find several peaks coming from different nuclei with similar abundances. Thus the average nucleus does not give a good description of the nuclear distribution. If the representative nucleus does coincide with the average of the distribution, it still would be necessary to include shell effects in the SNA. Contrary, at large temperatures the distributions can become very broad which cannot be taken into account in the SNA. In both cases the effect of the modified composition on capture rates and scattering amplitudes could be important. At $T=5$ MeV and $T=10$ MeV the shell effects become so weak, that the typical behavior of statistical models after the onset of the liquid-gas phase transition\footnote{To avoid misunderstandings: here we are refering to the binodal surface of bulk nuclear matter shown in Fig.~\ref{fig_bulkpd} and do not mean our Maxwell construction to uniform nuclear matter.} is recovered: the distributions change from steep exponentials to broad power-law and finally U-shape distributions with increasing density.  For such broad distributions the mass fractions of the different components change continuously. In general, the mass numbers increase with density. E.g.~at $Y_p=0.3$ and densities close to saturation all nuclei of the nuclear chart are populated up to the most heavy and super-heavy nuclei which exist in the mass table used. 

In the EOSs of Refs.~\cite{ lattimer91,shen98,shen98_2} only $\alpha$-particles are used to represent light clusters. At low temperatures $T\sim 1$ MeV this works well as the $\alpha$-particles give the main contribution. Then the $\alpha$-particle fraction is similar in the three different calculations. At high temperatures the whole ensemble of light nuclei becomes important, with a particular large contribution from deuterons, as also found in Refs.~\cite{horowitz06a,horowitz06b,oconnor07,arcones08,sumiyoshi08,roepke09,typel09}. The deuterons surmount the alphas in many cases. At densities slightly below the abundant appearance of heavy nuclei (or the transition to uniform nuclear matter) the whole ensemble of light clusters gives the main contribution to the composition. Then also differences in the $\alpha$-particle fractions of the NSE model compared to the SNA calculations occur and the $\alpha$-particle fraction differs significantly (one order of magnitude) from the total light cluster fraction of the NSE model. Here we find that for very low $Y_p$ the light clusters become very neutron rich and heavier compared to large $Y_p$. However, this might also occur as a compensation of the limited mass table which does not include heavy nuclei with very low $Z/A$. 

We compared the equation of state to the ones of LS and Shen et al. In general, up to densities $10^{-4}$ fm$^{-3}$ the three models agree very well. The differences of the NSE model to the two other EOSs are comparable to the differences between the LS and the Shen EOS. The NSE EOS is closer to the results of Shen et al., whose model is based on a similar RMF parameterization (TM1).  At low temperatures strong shell effects lead to narrow distributions in the NSE model so that the SNA is well applicable. In Refs.~\cite{blinnikov09} a comparison with the Shen EOS also showed the insensitivity of thermodynamic variables to the underlying physical description. The same result was obtained in Ref.~\cite{botvina08} in which the Statistical Multifragmentation Model (SMM) was compared to the LS EOS. However, at larger densities in our calculation some significant deviations occur. 

In our NSE model the shell effects are weakened at large $T$, and significant differences around the onset of the liquid-gas phase transition appear. This happens at $n_B>10^{-3}$ fm$^{-3}$ for $T=5$ MeV and $n_B>10^{-2}$ fm$^{-3}$ for $T=10$ MeV. The large contribution of light clusters causes an increased energy density (differences up to $\sim$10 MeV) and increased entropy per baryon (differences up to $\sim$1.5), resulting in a lower free energy density in a narrow density range. The pressure and chemical potentials also become slightly reduced. Compared to the SNA models, the heavy clusters appear at larger densities. With the onset of heavy clusters a sudden decrease in the entropy and energy densities is observed. In this density range, one to two magnitudes below saturation density, the baryon contribution to the total EOS is large, so that the mentioned differences are also significant for the total EOS. It is important to note that the two different SNA models do not show any remarkable differences here. We think that the formalism of Ref.~\cite{burrows84}, in which it was derived that the SNA does not influence the EOS, can not be applied in this density regime, as a Taylor-expansion around the representative nucleus is used, but the distributions are actually exponential or power-law. In a study of our NSE model without light nuclei with $A<20$ except for the alphas, we found that the differences became much weaker. Thus we conclude that it is the inclusion of light clusters (which cannot be represented by alpha-particles  in this density regime) which lead to the characteristic shape of the nuclear distributions and the appearance of heavy nuclei at larger densities. We expect that these effects cause the observed differences in the EOS compared to the two SNA models. In Ref.~\cite{ishizuka03} a noninteracting NSE model was compared to the Shen EOS and a bulk RMF EOS. They also found that the differences are most pronounced around the onset of the liquid-gas phase transition region. They came to the similar conclusion that at the borderline of the liquid-gas phase transition region (dubbed the boiling point by the authors) fluctuations are large and that it is necessary to take into account the fragment mass and isotope distribution. Other reasons for the observed differences could be the phenomenological character of our NSE model, the description of excited states of the nuclei, the different nuclear interactions, or missing medium effects. 

These results also do not take medium modifications beyond the excluded volume effects into account, which become relevant for the light clusters above $n_B=10^{-4}$ fm$^{-3}$ \cite{roepke09,typel09}. We stress, that in the LS and Shen et al.~EOSs no medium effects on the $\alpha$-particles are included either. However, we still achieve the dissolution of the light clusters close to saturation density by the excluded volume description. We did a comparison of the composition with detailed medium effects of Ref.~\cite{typel09} to our NSE model in which we included only the same light cluster up to $A\leq4$. For $T=10$ MeV and $Y_p=0.5$ we find that light clusters exist up to larger densities and appear with larger maximum fractions but the overall qualtitative behavior is similar. For $T=4$ MeV and $Y_p=0.5$ the differences are more pronounced, as our NSE model with excluded volume and interactions of the nucleons behaves similar to an ideal-gas NSE, almost until saturation density is reached. However, in the full calculation with all possible nuclei, for temperatures $T<10$ MeV and large densities heavy nuclei appear before the nucleon gas becomes very dense. A more detailed comparison of the NSE model with the works of \cite{roepke09,typel09} would be an interesting study. 

At larger densities the free energy of the NSE model becomes larger than in the EOSs of LS and Shen et al. We attribute this result to the limited description of the heavy nuclei due to the mass table used. Close to saturation density, the excluded volume effects become so strong that they can be seen directly in the EOS, e.g.~by an increased pressure. At slightly larger densities the uniform nuclear matter phase sets in, with the mixed phase given by a Maxwell construction. The chosen description of the phase transition to uniform nuclear matter leads to a continuous behavior of the free energy, the energy density and the entropy, the pressure and the total baryon chemical potential which are first derivatives of the free energy. The second derivatives behave discontinuously. We note that with the inclusion of trapped neutrinos in weak equilibrium the constant pressure plateau of the Maxwell construction is smeared out. 

\section{Outlook}
\label{sec_out}
Finally, we conclude that the NSE model presented here gives a good description of matter in supernovae across all relevant densities. Still, many aspects of the model can and have to be improved. First of all, a more elaborated description of the nontrivial temperature effects on the nuclei is wanted. Obviously, the approach presented is a very crude treatment of the temperature effects in nuclei. However, in Ref.~\cite{liu07} it was shown, that the NSE composition is rather insensitive to the nuclear partition function. The authors of Ref.~\cite{liu07} compared a NSE composition which is based on the ground-state partition function to the composition obtained with the detailed calculations of nuclear partition functions of Rauscher and Thielemann \cite{rauscher97,rauscher00,rauscher03}, and found only minor differences. Also in Ref.~\cite{blinnikov09} it was stated that the effect of nuclear excited states is not very large. However, we found in a tentative analysis that the inclusion of excited states of the light nuclei has a significant effect on the EOS at large temperatures, which has to be studied further. It would be very attractive to use the detailed internal partition functions in the tabular form of Rauscher \cite{rauscher97,rauscher00,rauscher03} and to compare the results with the simple semi-empirical degeneracy function used here. 

A mass table which includes also super-heavy nuclei, up to say $A\sim500$, is in principle necessary. With the mass table used here the model is too restrictive for very large densities where the heaviest nuclei appear. Also for low $Y_p<0.1$, where the nuclei become very neutron rich we are limited to nuclei above the neutron drip, restricting $Z/A$ from below. Furthermore, at such low $Y_p$ the free nucleon density becomes large, resulting in a possible modification of the nuclear binding energies which is not taken into account here. Also at intermediate temperatures $T \sim5$ MeV nuclei and unbound nucleons appear with similar abundance so that the influence of the unbound nucleons on the nuclear structure could be important. Due to the same reasons, in the case of cold deleptonized neutron stars where the proton fraction is very low, the composition of the NSE model is only robust at densities not much higher than the neutron drip density.

For the transition to uniform nuclear matter a Gibbs-construction based only on local charge neutrality is needed. Then the baryon density as well as the proton fraction could adjust in the non-uniform and the uniform nuclear matter phase. The phase transition would become smoother and the pressure would not be constant in the mixed phase any more. Furthermore, as the uniform nuclear matter phase can be interpreted as an infinitely large nucleus, we expect that with the use of the Gibbs-construction the restriction of the mass table would not play such an important role any more.

Because of the analytic formulation of the thermodynamic quantities it is possible to derive the second derivatives of the free energy of the NSE model, which are of great interest for applications of the EOS. These second derivatives, as e.g.~the heat capacity, can be expressed as simple functions of the three different components (nucleons, electrons and nuclei). Thus additional root findings are, if at all, only required for the evaluation of the second derivatives of the different components, which are rather easy to perform, but not for the system as a whole. We leave the derivation and discussion of the second derivatives for future studies.

The underlying assumptions in the NSE model for the excluded volume corrections which are used here are not the only possible ones. Currently the authors are exploring other possibilities which could give a more realistic description of the medium effects. Another major change would be the possibility to develop an extended model which combines the presented interacting NSE model with the Statistical Multifragmentation Model of Botvina and Mishustin \cite{bondorf95,botvina08,mishustin08}. This could improve the high density and high temperature behavior as it would allow to include weakening of shell effects at large temperatures or the appearance of super-heavy nuclei. 

The advantages of a NSE description are obvious. Most importantly, the SNA gives per se only an averaged approximation of the real composition and ignores shell effects. It would be interesting to further explore the consequences of the NSE model on electron and neutrino emission-, absorption- and scattering-reactions. Furthermore the NSE model is very fast, accurate and rather easy to handle. This allows to construct new EOS tables relatively quickly. Different parameterizations and models for the nuclear interactions as well as different mass tables can be implemented. The final aim of this work is to provide complete EOS tables, after the above limitations have been examined further, which then can be used in numerical simulations of core-collapse SN and other astrophysical systems. In addition to the EOS tables we will provide a computer program which calculates the full nuclear distributions. For exploratory studies an EOS table based on the current version of the NSE model is available from the authors upon request. It would be interesting to examine the consequences of a modified low- as well as a modified high-density behavior of the EOS within simulations. In addition, our simple EOS can be used as a reference for comparison with more elaborated models which focus on certain aspects of the SN EOS. 

The EOS derived in this work is not only of interest for supernova, but also for many other astrophysical scenarios in which similar thermodynamic conditions occur, as e.g.~neutron star mergers, collapsar model for gamma-ray bursts, accreting neutron stars or proto-neutron stars. The NSE model may especially be useful for the seed composition in nucleosynthesis calculations. The EOS of our statistical model directly contains the abundance patterns of the nuclei in a consistent way. The passing of matter through the liquid-gas phase transition region in which various nuclei are formed could be seen as a preprocess for the standard r-process nucleosynthesis process as noted in \cite{ishizuka03, botvina04}. The main characteristic features of the solar element abundances appear already in the element distributions of statistical models. Especially the formation of very heavy and super-heavy nuclei, as shown in Fig.~\ref{fig_ya}, is an interesting aspect by itself. Within supernova matter, especially because of the screened Coulomb forces of the nuclei and the large electron degeneracy, heavy nuclei are favored. If their vacuum-lifetimes were long enough they could possibly survive the supernova and could be present in cosmic rays, detectable here on earth. Besides the astrophysical context, our NSE model could be applied for low-energy heavy ion collisions for the description of statistical multifragmentation. It would be very attractive to compare our theoretical predictions with other existing models as e.g.~\cite{gross90, bondorf95} and experimental data.

\subsection*{Acknowledgments}
M.H.~acknowledges support from the Graduate Program for Hadron and Ion Research (GP-HIR). J.S.B.~is supported by the German Research Foundation (DFG) within the framework of the excellence initiative through the Heidelberg Graduate School of Fundamental Physics. This work has been supported by CompStar a research networking program of the European Science Foundation. We would like to thank T.~Fischer, I.~Mishustin, A.~Botvina, G.~R\"opke, and S.~Typel for stimulating discussions. 

\bibliographystyle{apsrev_nourl}
\bibliography{literat}
\end{document}